# PENERAPAN TEKNOLOGI PENGOLAH CITRA DIGITAL DAN KOMPUTASI PADA PENGUKURAN DAN PENGUJIAN BERBAGAI PARAMETER KAIN TENUN

Andrian Wijayono[1], Irwan[1], Siti Rohmah[1] & Valentinus Galih Vidia Putra[1]

Textile Engineering Departement, Politeknik STTT Bandung, Indonesia[1]

**Abstrak:** Kualitas merupakan salah satu hal yang penting untuk dijaga pada sebuah industri pertenunan. Seiring perkembangan zaman, perkembangan teknologi di bidang pengolahan citra dan komputasi telah mengubah metode lama evaluasi visual kain tenun menjadi lebih baik. Pada bab ini akan diterangkan mengeai beberapa implementasi teknik pengolah citra dan komputasi pada bahan kain tenun.

**Kata Kunci:** *kain tenun, pengolah citra digital.*

## 1. PENDAHULUAN

Kain tenun merupakan salah satu produk tekstil yang banyak diproduksi hingga saat ini, baik untuk keperluan tekstil sandang, maupun untuk tekstil non sandang. Untuk dapat memenuhi persyaratan mutu kain tenun yang diperlukan, maka proses pengendalian mutu yang harus dilakukan.

Seiring perkembangan pada dunia komputasi, perkembangan pada penerapan teknologi *image processing* pada bidang kain tenun banyak dilakukan. Dalam beberapa penelitian yang telah dilakukan, menunjukan bahwa dengan teknik *image processing* lebih mampu untuk melakukan proses evaluasi visual kain tenun yang lebih cepat dan lebih akurat. Pada bab ini akan dijelaskan beberapa penerapan teknik *image processing* pada pengendalian mutu kain tneun.

## 2. METODE PENGOLAHAN CITRA UNTUK MENGEVALUASI STRUKTUR ANYAMAN KAIN TENUN

Analisis visual dari sampel kain adalah proses penting untuk mereproduksi mengevaluasi karakteristik suatu kain. Pada dasarnya analisis ini mendefinisikan anyaman, kerapatan benang lusi dan benang pakan dan nomor benang lusi dan benang pakan dengan menggunakan mikroskop. Proses ini secara tradisional dilakukan oleh *inspector* manusia yang menggunakan alat kaca pembesar, penggaris dan beberapa alat sederhana lainnya untuk menghitung kerapatan dan secara visual menentukan pola tenun. Umumnya operasi manual seperti ini bersifat membosankan, menyita waktu dan merepotkan. Dengan demikian penilaian mungkin tidak konsisten atau cukup akurat karena dapat bervariasi dari satu *inspector* ke *inspector* yang lain.

Di sisi lain, perkembangan dinamis pada kemampuan komputer dan kapasitas penyimpanan membuka akses bagi analisis citra digital yang lebih maju untuk menggantikan operasi yang bergantung pada penglihatan manusia. Dengan menggunakan analisis citra digital memungkinkan analisis detail terhadap parameter struktur dasar produk tekstil [1]. Teknik tersebut digunakan sebelumnya untuk memperkirakan luas penampang serat wol [2]. Setelah itu, aplikasi lain muncul untuk memperkirakan penyimpangan campuran serat pada permukaan benang, untuk mengevaluasi kedewasaan kapas dan untuk menganalisis kerusakan serat wol [3 - 7]. Peneliti lain menggunakan analisis citra digital untuk mengkarakterisasi parameter struktur dasar permukaan benang seperti ketebalan, *hairiness* dan *twist* [1, 8, 9]. Analisis digital juga digunakan untuk mengkarakterisasi tekstur karpet selama pemakaian [10].

Teknik pengolahan citra dapat pula digunakan untuk menilai permukaan kain yang memiliki *pilling* dengan menganalisis kecerahan masing-masing *channel* (Merah, Hijau dan Biru) dari gambar kain berwarna. Teknik analitik digunakan untuk menetapkan nilai pilling untuk setiap sampel berdasarkan penghitungan *area pilled* yang diperoleh dari citra yang dianalisis [11]. Teknik pengolahan citra lainnya digunakan untuk mengukur *surface roughness* kain rajut. Gambar kain ditangkap melalui scanner beresolusi tinggi dan analisis analitik dilakukan untuk mendapatkan *index roughness* kain [12]. Kusut kain juga ditandai dengan memanfaatkan pengolahan citra melalui analisis ketinggian profil cahaya yang dibuat oleh kerutan kain. Parameter statistik untuk profil cahaya diperkirakan mengkarakterisasi kusut kain [13]. Transformasi frekuensi juga digunakan untuk memperkirakan fitur morfologi untuk jaringan *nonwoven* [14] dan untuk mengekstrak beberapa fitur gambar untuk mengklasifikasikan beberapa cacat kain rajutan [15]. Operasi koreksi, seperti perataan histogram dan erosi autokorelasi, juga digunakan pada aplikasi lain untuk mengklasifikasikan beberapa cacat kain tenun [16]. Filter Wiener digunakan untuk pengenalan pola tenun dengan cara mendekomposisi gambar kain menjadi dua gambar, yaitu gambar benang lusi dan gambar benang pakan [17]. Kemudian gambar lain diinisialisasi untuk menentukan garis grid yang mewakili sumbu tengah benang. Poin dimana sumbu sentral berpotongan didefinisikan sebagai titik *cross-over*. Teknik ini mengasumsikan bahwa benang lurus dan mengidentifikasi pola dengan memeriksa intensitas pada setiap titik silang. Pemeriksaan titik silang pada citra digital kain dilakukan hanya dengan memeriksa intensitasnya untuk memutuskan apakah titik tersebut dapat dikatakan sebagai *cross-over point* atau bukan.

Tujuan dari penelitian ini adalah untuk menggunakan analisis pengolahan citra untuk memperkirakan beberapa karakteristik struktural dari kain tenun dan untuk mengidentifikasi pola tenunan. Keberhasilan pendekatan pengolahan citra semacam itu akan memungkinkan analisis yang cepat dan akurat terhadap beberapa karakteristik struktur kain. Prosedur tradisional diketahui bersifat membosankan, menyita waktu dan merepotkan mata inspektur. Semua kekurangan ini akan dieliminasi saat prosedur tradisional digantikan oleh sistem komputer yang menangkap dan memproses gambar kain. Dalam karya ini, pendekatan pengolahan citra yang memanfaatkan filter Wiener disajikan untuk mengidentifikasi pola kain tenun dan memperkirakan beberapa karakteristik struktur kain. Enam kelompok sampel kain digunakan dalam pekerjaan ini termasuk tiga struktur kain yang berbeda, yaitu tenunan polos, kepar 3/1 dan satin 5, dengan masing-masing struktur mengandung dua konstruksi kain agar memiliki karakteristik struktur yang berbeda. Lima gambar diambil dari masing-masing kelompok sampel untuk dianalisis. Pola tenunan, kerapatan lusi dan pakan, serta diameter benang diidentifikasi dan dibandingkan dengan data sampel yang diperkirakan menggunakan prosedur manual tradisional.

Tiga struktur kain dipilih untuk penelitian ini dan masing-masing struktur kain diwakili oleh dua konstruksi kain. Ketiga struktur kain itu polos, twill 3/1 dan satin lima gun. Semua kain dibuat dari benang katun 100%. Keenam sampel kain diuji dengan menggunakan prosedur manual tradisional untuk mengidentifikasi struktur dan kerapatan kain di arah lusi maupun pakan. Jumlah benang setiap sampel diuji dan hasilnya sesuai dengan yang diperoleh dari produsen kain. Spesifikasi terperinci dari masing-masing sampel tercantum pada Tabel-1. Data mewakili nilai rata-rata yang diukur dengan kesalahan standarnya dan data disediakan oleh produsen. Semua sampel kain tidak berwarna kecuali sampel nomor empat, yang merupakan kain denim dengan benang lusi berwarna biru tua dan benang pakan berwarna putih.

Tabel-1 Spesifikasi sampel kain.

| ID | Fabric structure | Yarn density, thread per cm | | Yarn count, tex | |
|---|---|---|---|---|---|
| | | Warp | Weft | Warp | Weft |
| 1 | Plain 1/1 | 26 ± 0.409 | 35 ± 0.551 | 20 ± 1.194 | 28 ± 1.333 |
| 2 | Plain 1/1 | 30 ± 0.495 | 26 ± 0.491 | 34 ± 1.637 | 34 ± 1.543 |
| 3 | Twill 3/1 | 37 ± 0.562 | 20 ± 0.339 | 22 ± 1.301 | 16 ± 0.841 |
| 4 | Twill 3/1 | 26 ± 0.542 | 31 ± 0.486 | 14 ± 0.778 | 20 ± 1.109 |
| 5 | Satin 5 | 27 ± 0.530 | 18 ± 0.407 | 20 ± 1.261 | 14 ± 0.711 |
| 6 | Satin 5 | 57 ± 0.746 | 29 ± 0.542 | 46 ± 1.977 | 42 ± 2.044 |

Sebuah kamera CCD yang dilengkapi dengan lensa pembesar digunakan untuk menangkap gambar kain. Lima gambar berbeda diambil untuk setiap jenis sampel, yang didigitasi menggunakan *grabber frame (VGA Card)* dan dipindahkan ke komputer pribadi untuk disimpan. Ukuran gambarnya 512 × 512 piksel dengan resolusi 6500 piksel per inci. Semua gambar diproses menggunakan pemerataan histogram untuk memperbaiki tampilan visual. Gambar berwarna diubah menjadi gambar greyscale dua dimensi dengan 256 *grey level* untuk mempercepat proses pengolahan selanjutnya pada komputer. Sampel gambar untuk tiga struktur ditunjukkan pada Gambar-1 setelah konversi skala abu-abu

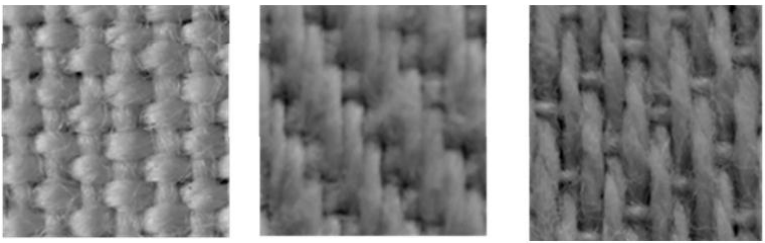

Gambar-1 Gambar *grey scale* untuk ketiga struktur kain; a) anyaman polos, b) anyaman keper, c) anyaman satin.

Filter Wiener diaplikasikan pada gambar kain *grey scale* untuk meregenerasi dua sub gambar dari gambar asli kainnya. Setiap sub-gambar hanya menampilkan satu kelompok dari dua kelompok benang dasar yang dikenal sebagai lusi dan pakan. Umumnya filter Wiener menggunakan konstanta spectrum warna untuk mengurangi *noise* di dalam garmbar. Filter Wiener dapat digunakan untuk menghitung nilai setiap piksel [17] dengan menggunakan persamaan berikut:

$$W(m,n) = \mu + \frac{S^2 - v^2}{S^2}[l(m,n) - \mu] \qquad (1)$$

$$\mu = \frac{1}{MN}\sum l(m,n) \qquad (2)$$

$$S^2 = \frac{1}{MN} \sum l^2(m,n) - \mu^2 \quad (3)$$

$v^2$ adalah nilai variansi dari *noise.*

Dimensi ($MxN$) dipilih berdasarkan aplikasi, dan metode filtrasi bergantung pada perhitungan statistik pada tetangga masing-masing piksel. Dengan asumsi bahwa matriks 2D mewakili gambar greyscale kain dan $m, n$ menunjukkan indeks piksel gambar, maka $l(m,n)$ akan menunjukkan intensitas piksel pada *grey level*, yang nilainya akan bervariasi dari 0 (hitam ), sampai 255 (putih).

Bentuk benang lusi atau pakan selalu berbentuk garis panjang. Oleh karena itu dengan menerapkan filter pada arah horizontal, tekstur vertikal dinetralkan dan sebaliknya. Jendela gambar dengan tinggi yang pendek dan lebar yang panjang akan menghasilkan sub-gambar yang hanya berisi kelompok benang pakan. Dalam hal ini ukuran jendela dipilih sebagai 5 × 60 piksel. Sebaliknya, jendela dengan tinggi dan lebar pendek akan menghasilkan sub-gambar yang hanya berisi kelompok benang pakan. Dalam hal ini ukuran jendela dipilih sebagai 60 × 5 piksel. Gambar 2 menunjukkan contoh gambar kain tenunan dalam bentuk *greyscale* yang sedang diselidiki serta sub-gambarnya yang dihasilkan dari penerapan filter Wiener. Proses pemerataan dan penyesuaian histogram diterapkan pada gambar yang dihasilkan untuk meningkatkan kualitas gambar, seperti yang ditunjukkan pada Gambar-2.

Beberapa *noise* dikenali di bagian atas dan bawah pada sub-gambar kelompok benang lusi dan juga pada sisi di sub-gambar kelompok benang pakan. Penghilangkan *noise* pada gambar akan mempermudah proses mendeteksi pinggiran benang dan akan memudahkan serta tidak mempengaruhi proses selanjutnya. Sub-gambar yang dihasilkan ditingkatkan lagi dengan memanfaatkan pemerataan histogram dan diubah menjadi gambar biner. *Clustering thresholding* atau metode Otsu digunakan untuk mendapatkan nilai ambang dari gambar kain untuk mengubahnya menjadi gambar biner [18]. Metode Otsu dianggap sebagai salah satu metode yang paling banyak dirujuk. Metode ini menetapkan ambang optimum dengan meminimalkan jumlah tertimbang dalam varians kelas untuk piksel latar depan dan latar belakang. Minimalisasi dalam varians kelas setara dengan maksimalisasi antara kelas yang tersebar. Hasil metode ini dianggap memuaskan bila jumlah piksel di setiap kelas saling berdekatan satu sama lain. Lubang kecil (3 × 3 piksel) dan garis pendek pendek yang muncul dalam gambar biner dianggap sebagai noise dan dihapus. Kemudian pinggiran masing-masing benang di kedua arah (lusi dan pakan) digariskan, seperti yang ditunjukkan pada Gambar 2.

**Perhitungan Tetal Benang dan Nomor Benang**

Garis tepi benang yang ditunjukkan pada langkah terakhir dari Gambar 2 selanjutnya digunakan untuk menghitung nilai rata-rata dari diameter benang di setiap arah lusi maupun pakan dengan menghubungkan resolusi gambar dengan jumlah piksel yang mewakili masing-masing lebar benang. Diameter rata-rata benang lusi dan benang pakan dapat dihitung nomor benangnya (Ne) dengan menggunakan hubungan berikut:

$$d \approx \frac{1}{28\sqrt{Ne}} \qquad (4)$$

Jumlah benang di setiap arah diidentifikasi dan digunakan untuk menghitung kerapatan benang di setiap arah dengan menggunakan informasi dari dimensi gambar. Teknik yang sama digunakan untuk menghitung jarak benang.

**Cover Factor Kain**

Secara umum, *cover factor* menunjukkan sejauh mana area kain ditutupi oleh satu kelompok benang, dua faktor penutup: satu untuk benang lusi dan yang lainnya untuk benang pakan. Pierce mempresentasikan persamaan berikut untuk menghitung faktor penutup untuk masing-masing arah, baik lusi maupun pakan [19]:

$$Cover\ factor = \frac{n}{\sqrt{N_e}} \qquad (5)$$

Dimana $n$ merupakan jumlah helai benang per inch kain, sedangkan $N_e$ adalah nomor benang dalam sistem penomoran inggris.

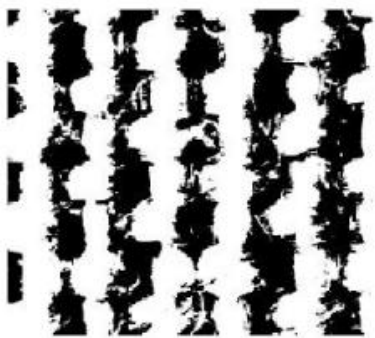
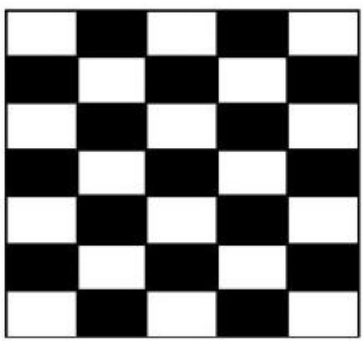

a)

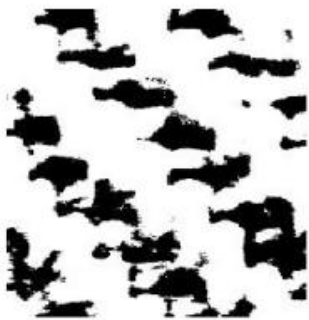
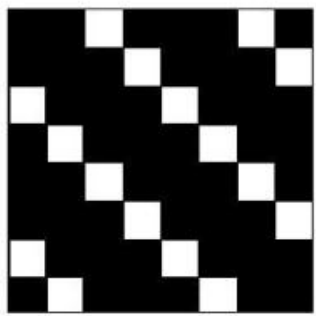

b)

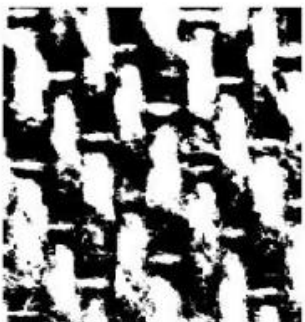
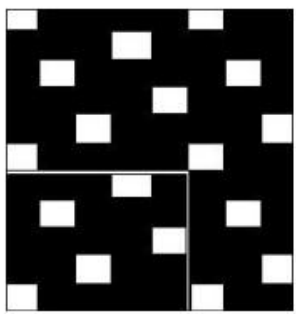

c)

*Original image converted into a binary image*     *Recognized structure*

Gambar-3 Citra *gray scale* dan hasil identifikasi struktur dari a) kain tenun polos, b) kain tenun keper dan c) kain tenun satin.

*Cover factor* mewakili kedua kelompok benang lusi dan pakan, yang didefinisikan sebagai luas total kain yang ditutupi oleh benang. Pendekatan sederhana digunakan untuk menghitung *fractional cover factor*, dengan mengasumsikan bahwa benang memiliki penampang melingkar. Jika diameter benang $d$ dan jarak antar benang adalah $s$, *fractional cover factor* dinyatakan sebagai $d/s$. Dalam model ideal, $s$ bernilai sama dengan $1/n$, dan karenanya *fractional cover*-nya adalah $d \times n$. Jika $C_W$ adalah penutup pecahan untuk lusi dan $C_F$ adalah untuk benang pakan, *cover factor* keseluruhan kain dapat dirumuskan sebagai berikut

$$Total\ Cover\ Factor = C_W + C_F - (C_W C_F) \tag{6}$$

### *Pattern Recognition* (Pengenalan Pola)

Garis tepi benang pada masing-masing sub-gambar diidentifikasi seperti yang disebutkan sebelumnya untuk setiap gambar kain, seperti yang ditunjukkan pada Gambar-2. Hanya garis besar benang yang ditangkap di setiap sub-gambar untuk menghasilkan dua gambar yang hanya berisi garis yang mewakili benang, dan dengan menambahkan kedua gambar ini, akan dihasilkan suatu gambar yang berisi garis grid. Setelah itu, silangan-silangan pada benang akan di identifikasi. Dengan menerapkan informasi ini pada gambar biner kain dengan menggunakan metode Otsu, daerah silangan akan didefinisikan dalam gambar biner. Kemudian intensitas semua piksel di dalam jendela dijumlahkan. Analisis menunjukkan bahwa daerah di mana benang pakan menyilang diatas benang lusi lebih banyak dibandingkan dengan benang pakan menyilang dibawah benang lusi. Untuk tujuan ini, struktur kain diidentifikasi seperti yang ditunjukkan pada Gambar-3, di mana tanda putih bermakna daerah benang lusi menyilang di atas benang pakan. Pendekatan ini mengevaluasi pola tenun berdasarkan intensitas di setiap area silangan, yang membuatnya lebih andal dan mampu mendeteksi berbagai pola tenun.

Penerapan pengolahan citra telah diterapkan pada berbagai gambar kain untuk menghitung diameter benang, mengidentifikasi pola tenunan serta untuk mengukur parameter lainnya. Penerapan pengolahan citra diterapkan untuk semua gambar dan hasil pengenalan pola dibandingkan dengan pola tenun yang diketahui. Pendekatan yang disajikan mampu mengidentifikasi pola struktur kain dengan benar untuk semua sampel kecuali untuk sampel nomor 4 (kain denim). Masalah utama dengan sampel tersebut adalah perbedaan warna antara benang lusi dan benang pakan. Warna benang lusi berwarna biru tua (terlalu gelap) dan benang pakan berwarna putih (terlalu terang). Proses dekomposisi pada kain tersebut dapat dikatakan gagal. Karena adanya perbedaan warna antara benang lusi dan pakan, teknik pengolahan citra tidak dapat mengenali batas benang dan karenanya tidak ada daerah silangan untuk sampel tersebut. Gambar-4 dan Gambar-5 merupakan grafik yang menunjukkan hasil untuk tetal benang dan nomor benang dibandingkan dengan nilai yang diukur melalui prosedur tradisional. Hasil menunjukkan kesepakatan yang baik antara kedua prosedur tersebut. Beberapa perbedaan antara hitungan yang dihasilkan dari pendekatan dan hitungan yang diukur diidentifikasi untuk kain anyaman keper dan kain anyaman satin, alasannya didasarkan pada konsep mendapatkan hitungan dari pendekatan gambar. Pendekatan gambar menghitung lebar proyeksi benang, bukan diameter benang. Hasil pengukuran proyeksi benang tersebut yang kemudian digunakan untuk menghitung nomor benang. Terdapat perbedaan yang signifikan dalam nomor benang pakan terhadap nomor benang lusi, baik pada kain anyaman keper dan satin. Benang pakan memiliki tegangan yang lebih sedikit dibandingkan benang lusi selama proses pertenunan, hal tersebut memungkinkan benang pakan menjadi lebih pipih, terutama bila mereka memiliki ruang, yang disediakan dalam kain dengan tetal yang rendah dan / atau struktur kain yang memiliki Panjang *floating* benang yang relatif panjang, seperti anyaman keper dan satin. Ini menjelaskan mengapa hampir tidak ada perbedaan antara nomor pakan untuk tenunan polos, namun perbedaan tersebut terlihat jelas pada anyaman satin dan keper.

Hasil yang tercantum pada Tabel-3 menunjukkan jarak antar benang yang dihitung dengan pendekatan pengolahan citra digital. Gambar-6 menunjukkan perbandingan hasil pendekatan pengolahan citra untuk *fabric cover* kain dengan nilai estimasi, yang dihitung dari data yang diukur, yaitu tetal kain dan nomor benang. Berdasarkan hasil tersebut, terlihat bahwa tidak ada banyak perbedaan yang sangat jauh antara hasil nilai pengukuran dan hasil estimasi (tidak termasuk hasil sampel no 4). Berdasarkan hasil penelitian, hasil pengolahan citra pada parameter tetal kain, proyeksi diameter benang, jarak benang dan *fabric cover* kain, pengolahan citra dapat memberikan hasil cenderung lebih akurat. Akan tetapi, terdapat variasi dalam hasil perhitungan benang untuk struktur kain yang memiliki *floating* panjang karena adanya pemipihan benang, yang mempengaruhi diameter benang yang diproyeksikan. Oleh karena itu dapat disimpulkan bahwa teknik pengolahan citra mampu menganalisa kain yang memiliki benang lusi dan benang pakan dengan warna yang sama atau kain yang berwarna polos. Penerapan pengolahan citra mampu mengukur tetal benang lusi dan benang pakan, jarak benang, diameter benang lusi dan benang pakan dan nomor benang lusi dan benang pakan. Selain itu, pendekatan yang dikembangkan berhasil mengidentifikasi pola tenun yang berbeda. Dapat diprediksi bahwa pendekatan yang dikembangkan akan dapat mengidentifikasi berbagai macam pola setelah dapat mengidentifikasi batas-batas benang.

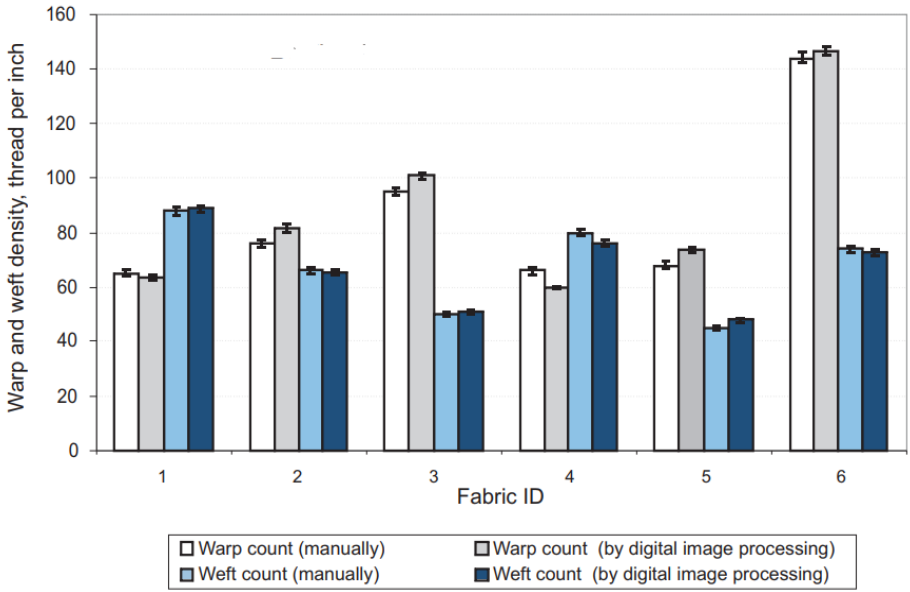

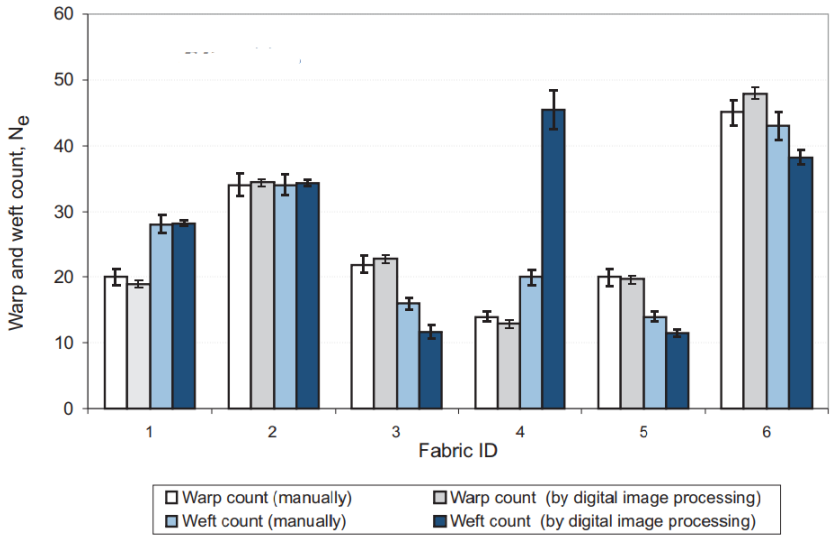

Gambar-4 Perbandingan hasil pengukuran tetal lusi dan tetal pakan menggunakan cara *image processing* dan cara manual

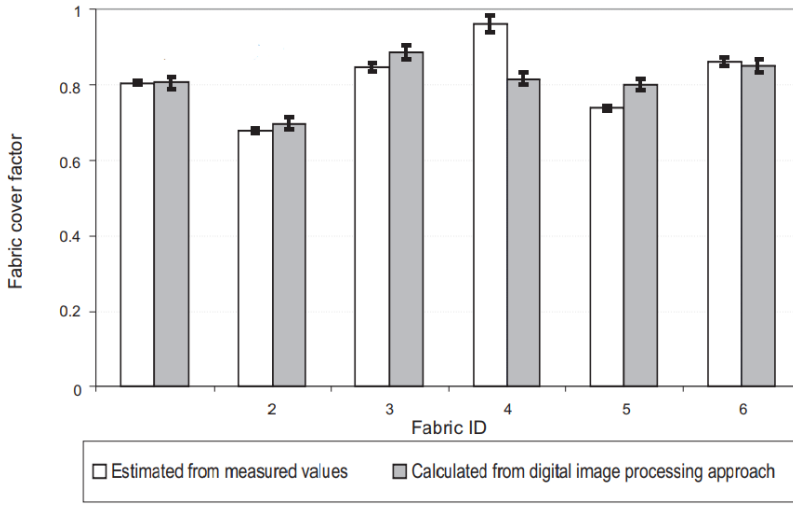

Gambar-5 Perbandingan nomor benang lusi dan nomor benang pakan menggunakan cara *image processing* dan cara manual

Gambar-6 Perbandingan *fabric cover* menggunakan cara *image processing* dan cara manual

Tabel-2 Hasil pengukuran pengolah citra digital pada nilai rata-rata diameter benang dan rata-rata *yarn spacing* yang diperoleh dari metoda *image processing* dan perhitungan

| ID | Mean yarn diameter, mm | | Mean calculated yarn spacing, mm | |
|---|---|---|---|---|
| | Warp | Weft | Warp | Weft |
| 1 | 0.208 | 0.170 | 0.373 | 0.356 |
| 2 | 0.155 | 0.155 | 0.432 | 0.406 |
| 3 | 0.191 | 0.239 | 0.254 | 0.533 |
| 4 | 0.254 | 0.191 | 0.406 | 0.432 |
| 5 | 0.206 | 0.241 | 0.457 | 0.559 |
| 6 | 0.132 | 0.147 | 0.178 | 0.432 |

Penelitian ini difokuskan pada identifikasi pola struktur anyaman selain untuk mengevaluasi karakteristik lainnya dengan memanfaatkan pendekatan pengolahan citra digital. Pendekatan yang dikembangkan menggunakan filter Wiener untuk menguraikan gambar kain menjadi dua sub gambar, yang masing-masing berisi kelompok benang lusi dan kelompok benang pakan. Sub-gambar selanjutnya dianalisis untuk menguraikan batas-batas benang. Diameter benang, jarak benang, jumlah benang, kepadatan di kedua arah dan faktor penutup kain telah diukur pada penelitian ini. Batas benang tadi selanjutnya digunakan untuk mengidentifikasi daerah silangan atau *cross-over area*, yang kemudian diproses untuk mengidentifikasi struktur kain. Enam sampel kain digunakan dalam penelitian ini untuk mengevaluasi pendekatan yang dikembangkan. Sampel mencakup tiga struktur kain dengan dua konstruksi untuk masing-masing struktur. Sampel dianalisis secara manual dengan menggunakan kaca pembesar dan hasilnya dibandingkan dengan pendekatan yang dikembangkan. Hasil pendekatan menunjukkan hasil yang mendekati untuk sampel yang memiliki warna yang sama untuk benang lusi dan pakan. Di sisi lain, pendekatan yang dikembangkan gagal menganalisis sampel yang memiliki perbedaan warna benang lusi dan benang pakan yang ekstrem. Pendekatan ini tidak dapat mengidentifikasi batasan benang dari sampel ini. Variasi besar dalam warna benang lusi dan benang pakan dalam sampel ini menambah kesulitan pada proses pendekatan pengolahan gambar dan memberikan hasil yang tidak valid. Pendekatan yang dikembangkan juga memberi kita pemahaman yang lebih baik tentang bagaimana proses tenun bisa mengubah beberapa dimensi benang, sehingga memberikan hasil yang lebih akurat untuk hasil pengukuran jarak benang dan *fabric cover*.

## 3. PENGUKURAN TETAL KAIN DENGAN MENGGUNAKAN METODE FOURIER TRANSFORM

Tetal benang pada kain tenun merupakan salah satu parameter yang penting pada industri pertenunan, yang biasanya diukur dengan cara dekomposisi kain secara manual. Metode yang digunakan untuk menentukan tetal tersebut yaitu dengan cara menghitung jumlah benang lusi dan benang pakan pada suatu unit panjang pada kain dengan menggunakan lup atau menggunakan kaca pembesar. Metode ini dianggap sebagai suatu metode yang bersifat *time comsuming* dan membutuhkan ketelitian yang lebih, tetapi tidak dapat menutup kemungkinan adanya kesalahan pada saat pengukuran tetal akibat kelelahan mata operator yang mengukur tetal dalam waktu yang lama. Berdasarkan hal tersebut, telah dikembangkan suatu sistem otomatis yang dapat mengukur tetal benang pada kain tenun.

Analisis citra telah terbukti sebagai salah satu metode yang efisien untuk menganalisis kerapatan benang pada kain tenun. Pada pengukuran tetal menggunakan analisis citra, ada dua hal yang dilakukan, yaitu menghitung jumlah benang pada kain tenun dan mengukur dimensi fisik dari citra kain. Secara umum, pengukuran dimensi fisik dari citra kain tersebut sudah termasuk kedalam sistem akuisisi gambar. Pan dkk (2015) telah melakukan suatu penelitian untuk mengukur jumlah benang pada suatu kain tenun dengan menggunakan suatu metode analisis citra digital yang terdiri atas dua proses, yaitu dengan proses metode analisis *frequency-domain* dan metode analisis *time-domain*.

Gambar kain tenun adalah tekstur periode khas yang dibentuk oleh benang lusi dan benang pakan. Hal ini dapat diubah menjadi domain frekuensi dengan transformasi Fourier. Puncak yang mewakili frekuensi elemen periodik terletak pada spektrum untuk menghitung kerapatan kain [1 - 7]. Puncak akan sesuai dengan informasi berkala atas susunan lusi dan susunan pakan yang digunakan untuk merekonstruksi gambar kain tenun. Gambar yang direkonstruksi dan hanya berisi benang lusi atau benang pakan digunakan untuk mengenali kerapatan kain tenun [8 - 10]. Beberapa periset telah mencoba menggunakan transformasi wavelet untuk merekonstruksi gambar kain di sepanjang arah lusi dan pakan dan yang digunakan untuk menghitung kerapatan kain tenun [11-13].

Dalam citra kain tenun, nilai abu-abu *pixel* dalam benang lebih besar dari nilai abu-abu piksel di celah benang. Dengan proyeksi abu-abu, benang lusi dan benang pakan pada gambar dapat ditemukan [14 - 19]. Dalam gambar kain yang ditangkap, celah dari benang dapat ditemukan dengan menggunakan metode *greyprojection* [20]. Matriks koherensi abu-abu juga digunakan untuk mengenali kerapatan kain tenunan [21].

Metode untuk memeriksa kerapatan kain tenun telah dijelaskan dalam banyak referensi. Terdapat banyak kesulitan dalam pengukuran kerepatan dengan cara otomatis, yaitu dapat menimbulkan kesukaran dalam analisis selanjutnya, seperti ketidakrataan benang, kelengkungan benang lusi dan benang pakan, dan bulu dari benang.

**Metode Penangkapan Gambar**

Tujuan dari penelitian yang dilakukan Pan dkk (2015) adalah untuk mendapatkan sistem inspeksi *real-time* untuk mengukur kerapatan kain tenun. Pemindai tipe *scanner* dapat menangkap gambar kain dengan resolusi tinggi, namun model akuisisi tersebut tidak sesuai untuk pemrosesan *real-time* [22]. Sebuah kamera CCD dapat digunakan untuk membangun sistem pengolahan citra real-time. Namun pembesaran lensa sederhana dari kamera CCD tidak cukup besar untuk membedakan benang pada kain. Mikroskop digital dipilih sebagai perangkat keras untuk akuisisi gambar. Pembesaran mikroskop yang digunakan adalah 10 - 40 dan resolusi tertinggi kamera CCD pada mikroskop digital adalah sekitar tiga mega *pixel*. Dalam sistem perangkat lunak, saat mengambil gambar kain dengan mikroskop digital, tiga jenis resolusi disediakan untuk digunakan, yaitu 512 × 384 piksel, 1024 × 768 piksel, dan 2048 × 1536 piksel. Pembesaran *default* adalah 10 dan resolusi *default* 512 × 384 piksel dalam sistem inspeksi. Resolusi gambar akan ditentukan berdasarkan parameter struktur tenunan. Jika benang tidak dapat dilihat dengan jelas dengan pembesaran dan resolusi *default*, maka perbesaran dan resolusi yang lebih besar akan digunakan. Struktur sistem pengukuran kerapatan benang dari kain tenun dalam percobaan ditunjukkan pada Gambar-7. Terdapat sebuah sumber cahaya digunakan untuk menerangi sampel kain. Kabel data USB digunakan untuk mentransfer data gambar dari kamera digital ke komputer dan kerapatan benang dari sampel kain tenun dapat diperiksa dengan perangkat lunak.

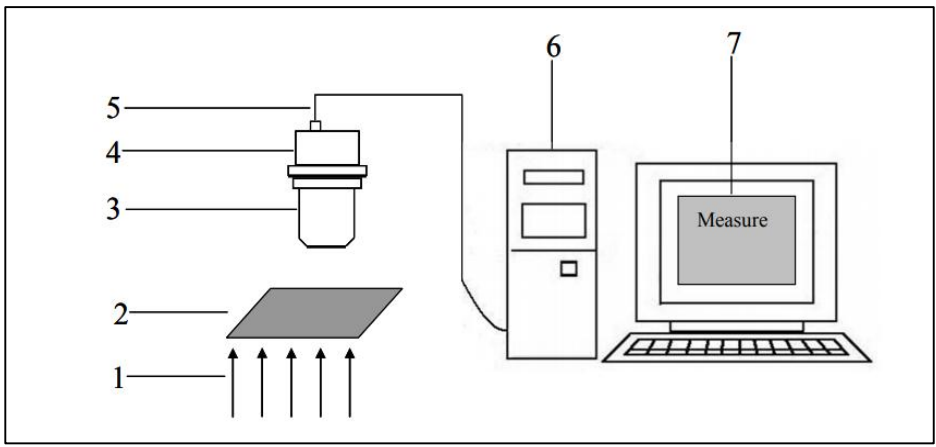

Gambar-7 Sistem pengukuran kerapatan kain tenun, (1) cahaya yang ditransmisikan, (2) sampel kain, mikroskop (3), (4) kamera digital, kabel data (5) usb, (6) komputer, (7) perangkat lunak

Ukuran sampel kain yang digunakan untuk memeriksa kerapatan (tetal) tidak lebih besar dari 10 x 10 cm. Sampel kain diregangkan di sepanjang arah yang berbeda. Benang lusi kemungkinan tidak tegak lurus terhadap benang pakan, oleh karena itu kerapatan lusi dan pakan diperiksa secara terpisah pada sistem pengukuran. Kelompok benang yang harus diukur diletakkan pada arah dengan lebih banyak *pixel*. Di dalam sistem pengukuran, angka defaultnya adalah 512 piksel dengan ukuran fisik sebenarnya 1,2 cm. Pembesaran mikroskop digital yang sebenarnya didapatkan melalui proses kalibrasi

Jika kerapatan kain adalah 100 threds / cm, satu benang dan silangan akan menempati sekitar 4,3 piksel dengan resolusi *default*. Resolusinya cukup untuk membedakan setiap benang pada kain. Jika nilai kerapatan benang lebih besar dari ini, operator dapat memilih resolusi yang lebih besar pada perangkat lunak atau menyesuaikan pembesaran mikroskop untuk menangkap gambar kain dengan resolusi lebih tinggi.

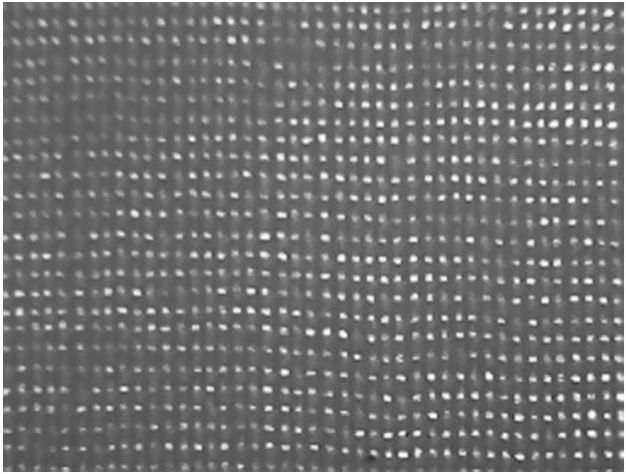

Gambar-8 Citra digital kain polos

Teradapat dua jenis model pencahayaan yang dapat digunakan untuk memeriksa kerapatan benang dari kain tenun pada sistem, yaitu dengan sistem ditransmisikan dan dipantulkan. Model pencahayaan yang digunakan pada penelitian Pan dkk (2015) merupakan sistem ditransmisikan, dimana identifikasi dilakukan berdasarkan cahaya yang ditransmisikan melalui celah benang. Nilai abu-abu dari piksel pada titik silang benang akan lebih besar dari nilai abu-abu piksel pada benang (non-silangan). Gambar-8 menunjukkan gambar kain anyaman polos yang ditangkap oleh sistem yang diusulkan dengan parameter *default*.

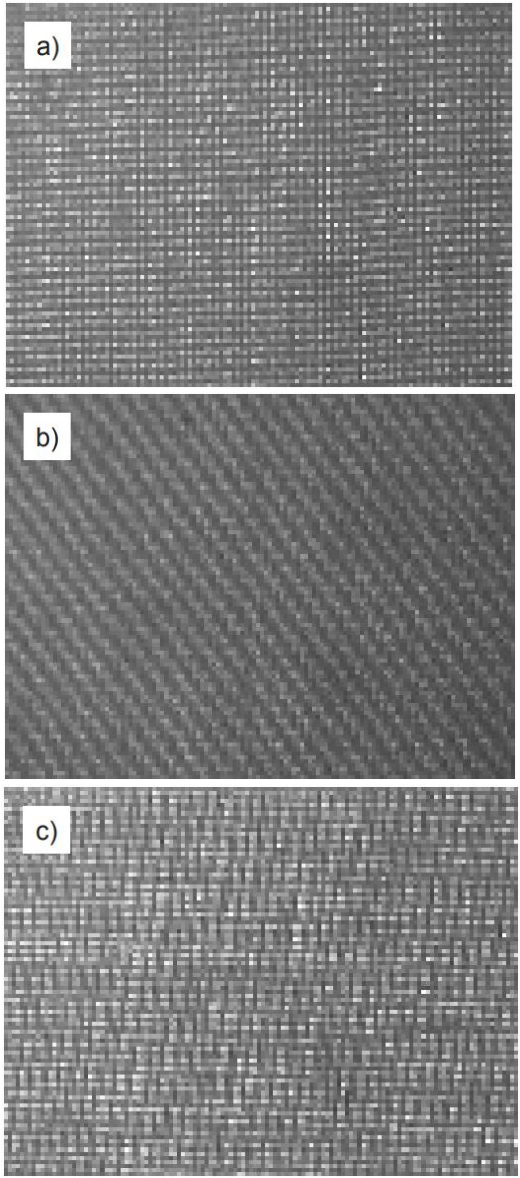

Gambar-9 Kenampakan kain yang digunakan dengan konstruksi polos 14,6 tex x 14,6 tex (a), keper 18,2 tex x 27,8 tex (b) dan satin 7,3 tex x 7,3 tex (c)

Tiga pola dasar kain polos, kain twill dan kain satin telah digunakan untuk menguji keefektifan sistem tersebut. Kain katun dengan jumlah benang dan rentang kerapatan yang berbeda dipilih untuk menguji kestabilan hasil sistem pendeteksian dengan metode ini. Tujuh kain polos (P1 - P7), lima kain keper (T1 - T5) dan tiga kain satin (S1 - S3) digunakan dalam percobaan. Gambar-9 menunjukkan kenampakan dari masing-masing kain tersebut.

Benang lusi dan benang pakan disilangkan dalam kain dan tidak dapat dihitung dalam satuan panjang untuk menghitung kerapatan dengan metode pengukuran gambar sederhana. Benang lusi dan benang pakan harus dipisahkan dengan analisis citra sebelum dapat diperiksa kerapatannya.

Sinyal periode yang berbeda dari citra kain dicampur bersama dalam suatu wadah yang bernama *timedomain*, sehingga sulit untuk digunakan untuk menganalisis tekstur secara langsung. Dalam domain frekuensi, periode susunan benang dapat dengan mudah dianalisis. Oleh karena itu, dalam sistem pengukuran ini, transformasi Fourier dilakukan pada citra kain dan kerapatan benang dapat dianalisis dalam domain frekuensi Fourier.

Adanya sinyal bising yang mengganggu citra kain secara periodik, seperti pencahayaan yang tidak merata dan bulu pada benang (*hairiness*). Gambar yang direkonstruksi dengan beberapa puncak di domain frekuensi [1 - 10] akan mengabaikan terlalu banyak sinyal detail pada gambar asli dan benang tidak akan tersegmentasi dengan benar jika puncaknya tidak berada pada posisi yang benar. Reabilitas algoritma yang sudah ada saat ini untuk mendeteksi kerapatan belum dapat memenuhi kebutuhan sistem pemeriksaan. Untuk mengatasi masalah di atas, daerah (tidak hanya beberapa puncak) yang sesuai dengan periodik benang lusi dan benang pakan dalam spektrum amplitudo transformasi Fourier dipilih untuk merekonstruksi citra kain. Kerapatan benang benang lusi dan benang pakan diperiksa pada gambar yang direkonstruksi.

Sebagai array diskrit dua dimensi dapat dilihat pada gambar skala abu-abu, transformasi Fourier dua dimensi (2D) dipilih untuk menganalisis citra kain tenunan. Gambar kain dan hasil transformasi Fourier didefinisikan sebagai f (x, y) dan F (u, v), di mana x, y adalah variabel koordinat dalam domain waktu dan u, v adalah variabel koordinat dalam domain frekuensi. Spektrum amplitudo M (u, v) dari transformasi Fourier dapat dihitung sebagai berikut

$$M(u,v) = \sqrt{R^2(u,v) + I^2(u,v)} \tag{7}$$

dimana, $R(u,v)$ adalah bagian nyata dan $I(u,v)$ adalah bagian imaginer.

Untuk mengamati hasil transformasi Fourier dengan mudah, spektrum amplitudo $M(u,v)$ ditunjukkan sebagai gambar abu-abu. Rentang numerik $M(u,v)$ terlalu besar untuk menggunakan transformasi linier sederhana untuk mengubah $M(u,v)$ ke dalam kisaran nilai skala gambar abu-abu (0 - 255).

Oleh karena itu, transformasi Log digunakan untuk mempersempit rentang numerik $M(u,v)$ untuk menunjukkannya sebagai gambar. Nilai "1" ditambahkan ke spektrum amplitudo untuk memastikan bahwa hasil fungsi log adalah bilangan positif. Spektrum amplitudo setelah transformasi Log didefinisikan sebagai $StreM(u,v)$, seperti pada persamaan berikut

$$StreM(u,v) = log_2[M(u,v) + 1] \tag{8}$$

Setelah mempersempit rentang numerik, transformasi linier digunakan untuk mengatur nilai amplitudo ke kisaran [0, 255], seperti yang ditunjukkan pada persamaan berikut ini

$$TM(u,v) = floor\left[\frac{StreM(u,v) - T_{min}}{T_{max} - T_{min}} x255 + 0{,}5\right] \tag{9}$$

dimana, TM (u, v) adalah spektrum amplitudo setelah transformasi linier; Tmax dan Tmin mewakili nilai maksimum dan minimum StreM (u, v), dan floor () mewakili fungsi rounded down.

Puncak yang sesuai dengan informasi berkala dari kain berada di daerah dengan frekuensi rendah, yaitu berada di garis batas spektrum amplitudo. Dengan demikian pusat gambar kemudian didefinisikan sebagai asal koordinat, dan sumbu x dan sumbu y berada disepanjang arah horizontal dan vertikal di sistem inspeksi. Kuadran pertama dan ketiga, dan kuadran kedua dan keempat dipertukarkan pada masing-masing kuadran, hal tersebut disebut sebagai pergeseran frekuensi. Frekuensi rendah kemudian bergeser ke tengah dan frekuensi tinggi masuk ke batas spektrum amplitudo.

Hasil transformasi Fourier terbalik digunakan untuk merekonstruksi citra kain. Nilai amplitudo dari invers transform diperoleh sebagai $M(u,v) = \sqrt{R^2(u,v) + I^2(u,v)}$ dan kemudian diubah menjadi [0, 255] dengan persamaan $StreM(u,v) = log_2[M(u,v) + 1]$ dan $TM(u,v) = floor\left[\frac{StreM(u,v) - T_{min}}{T_{max} - T_{min}} x255 + 0{,}5\right]$. Pada gambar yang direkonstruksi, citra benang pakan dikeluarkan sebagai suatu output, sedangkan benang lusi tidak dikeluarkan, atau sebaliknya. Setelah itu, beberapa metode analisis citra digunakan untuk menemukan benang dan menghitung jumlah benang. Berdasarkan jumlah benang yang ditemukan, kerapatan benang pada kain selanjutnya ditentukan.

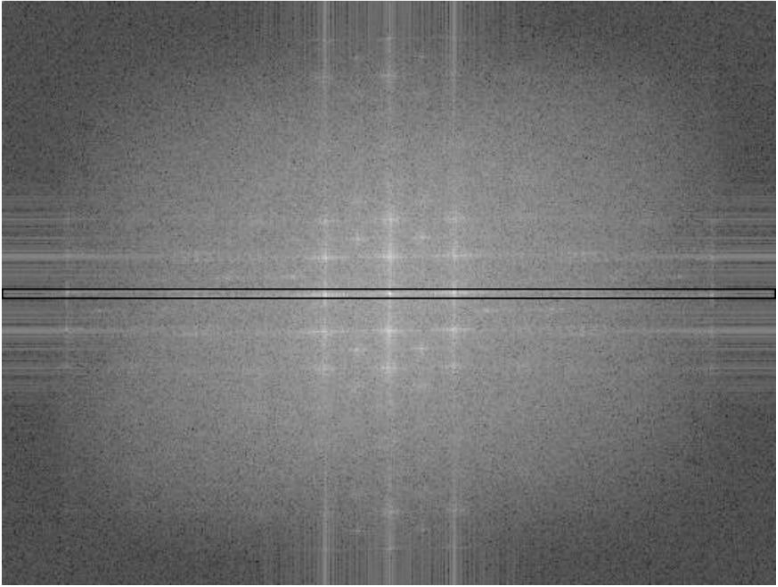

Gambar-10 Gambar spektrum amplitudo dari kain

Transformasi FFT (fast fourier transform) 2D digunakan untuk mengolah gambar kain tenunan, yang ditunjukkan pada Gambar-8, dan spektrum amplitudo kemudian diperoleh seperti yang ditunjukkan pada Gambar-10. Jumlah benang pada gambar kain dapat dihitung dengan hubungan antara puncak dan komponen arah [23], dan kemudian kerapatan lungsin dan pakan dapat diperiksa; Tapi sulit untuk membedakan benang-benang itu di dalam batas gambar. Kepadatan pada metode ini tidak bisa diperiksa dengan presisi tinggi. Area fitur yang berisi puncak dipilih untuk merekonstruksi gambar kain dengan benang lusi atau benang pakan. Sinyal periodik dari benang lusi diwakili sebagai garis terang horizontal di sepanjang pusat spektrum amplitudo dan sinyal periodik dari benang pakan diwakili sebagai garis terang vertikal di sepanjang pusat. Daerah yang berisi garis terang horizontal atau vertikal dipilih untuk merekonstruksi benang lusi atau benang pakan dengan sinyal frekuensi.

***Image segmentation***

Metode proyeksi abu-abu dapat digunakan untuk menemukan benang lusi atau benang pakan, namun mudah dipengaruhi oleh keadaan lurus tidaknya benang pada kain. Dalam gambar yang direkonstruksi, ada perbedaan signifikan dalam nilai abu-abu benang dan celah benang. Metode segmentasi *threshold* digunakan untuk memisahkan benang dari latar belakang.

Saat menangkap gambar kain, sumber cahaya digunakan untuk menerangi kain. Hal ini menyebabkan adanya kemungkinan perbedaan kecerahan antara daerah tengah dan daerah marjinal pada gambar. Sulit untuk mensegmentasikan semua benang pada gambar kain dengan metode *threshold* secara menyeluruh. Untuk menghilangkan ketidakrataan pencahayaan, metode *thresholding* adaptif lokal digunakan untuk mendegmentasikan benang lusi atau benang pakan pada gambar kain yang direkonstruksi.

Ada dua faktor utama untuk melakukan metode *threshold* adaptif local, salah satunya adalah ukuran jendela lokal dan metode *threshold* yang digunakan di jendela lokal. Pilihan ukuran jendela relatif terhadap ukuran diameter benang tunggal. Jendela lokal harus mencakup benang dan celah. Ukuran jendela sebaiknya paling tidak lebih besar dari diameter benang. Nomor *pixel* dalam benang ditentukan oleh pembesaran. Dalam percobaan, 32 × 32 *pixel* diatur sebagai ukuran jendela *default* lokal. Ukuran jendela lokal akan berubah apabila dilakukan pembesaran gambar kain.

Metode ambang niblack [22] dipilih untuk memproses *pixel* di jendela lokal. Nilai ambang dapat diperoleh sebagai berikut

$$T = u + k\,\sigma \qquad (10)$$

dimana, $u$ adalah nilai rata-rata piksel di jendela lokal dan $\sigma$ adalah nilai deviasi standar piksel di jendela lokal; $k$ adalah nilai konstan yang akan ditentukan dalam eksperimen.

Gambar kain yang ditangkap adalah semacam gambar yang dipancarkan, sehingga nilai abu-abu piksel pada benang lebih rendah daripada pada daerah silangan benang. Dalam proses ambang batas, piksel dengan nilai abu-abu lebih besar dari nilai ambang batas T berada di daerah silangan benang dan piksel lainnya ada di dalam benang.

Rekonstruksi benang lusi diambil sebagai contoh untuk menggambarkan prosedurnya. Saringan pita horizontal di sekitar garis terang horisontal dipilih untuk menyaring sinyal frekuensi. Sinyal yang disaring digunakan untuk merekonstruksi gambar kain dengan metode transformasi Fourier terbalik.

Untuk memastikan puncak yang sesuai dengan periodik benang lusi termasuk ke wilayah yang dipilih, sinyal frekuensi lebih kecil dari tiga piksel semuanya dipilih untuk membentuk pita penyaring frekuensi yang direkonstruksi. Template filter $TP(u,v)$ yang mengandung 0 dan 1 dibentuk sesuai dengan ukuran gambar yang diambil saat merancang sistem. Untuk mempertahankan sinyal frekuensi, nilai posisi yang sama pada template ditetapkan sebagai 1. Jika tidak, nilai piksel pada template ditetapkan sebagai 0. Saat membuat template, garis horizontal di sepanjang komponen langsung dan garis horizontal Mengambang tiga piksel ke atas dan ke bawah dipilih sebagai daerah penyaringan.

Hasil transformasi Fourier diberi *dotmultiplied* dengan template, dan kemudian pita frekuensi fitur $InvF(u,v)$ untuk merekonstruksi citra benang lusi dapat diperoleh seperti yang ditunjukkan pada persamaan berikut

$$InvF(u,v) = F(u,v) x TP(u,v) \qquad (11)$$

Spektrum amplitudo yang disaring ditunjukkan sebagai persegi panjang yang diberi label dengan garis hitam pada Gambar-10. Dari gambar tersebut, dapat dilihat bahwa puncak yang sesuai dengan periodik benang lusi termasuk dalam pita frekuensi yang dipilih. Citra dapat direkonstruksi dengan sinyal frekuensi hasil penyaringan dan benang lusi yang dipisahkan dari benang pakan bisa diperoleh. Hasil rekonstruksi benang lusi diperoleh seperti yang ditunjukkan pada Gambar-10. Dari gambar tersebut dapat dilihat bahwa sinyal benang pakan dikeluarkan saat sinyal benang lungsin dipertahankan. Dalam gambar ini, jumlah benang lusi dapat dengan mudah dihitung. Posisi dan keadaan benang lungsin pada hasil rekonstruksi masih sama seperti pada gambar aslinya, yang ditunjukkan pada Gambar-8. Intensitas cahaya benang pada Gambar-11 sama seperti pada gambar aslinya. Daerah terang pada gambar mewakili celah antar benang, sedangkan daerah gelap mewakili benang lusi.

Setelah mendapatkan citra benang lusi yang direkonstruksi, beberapa metode pengolahan citra dapat digunakan untuk menghitung jumlah benang lusi untuk mengukur kerapatan lusi. Posisi lusi harus ditempatkan sebelum menghitung benang. Karena nilai abu-abu antara benang dan celah jauh berbeda secara signifikan, metode ambang batas digunakan untuk memproses gambar yang direkonstruksi untuk menemukan benang dan celah benang. Metode *threshold* Niblack lokal digunakan untuk mengukur citra untuk menemukan benang. Nilai optimal k dipilih sebagai 0,2 dengan banyak percobaan. Hasil *threshold* tersebut ditunjukkan pada Gambar-12.

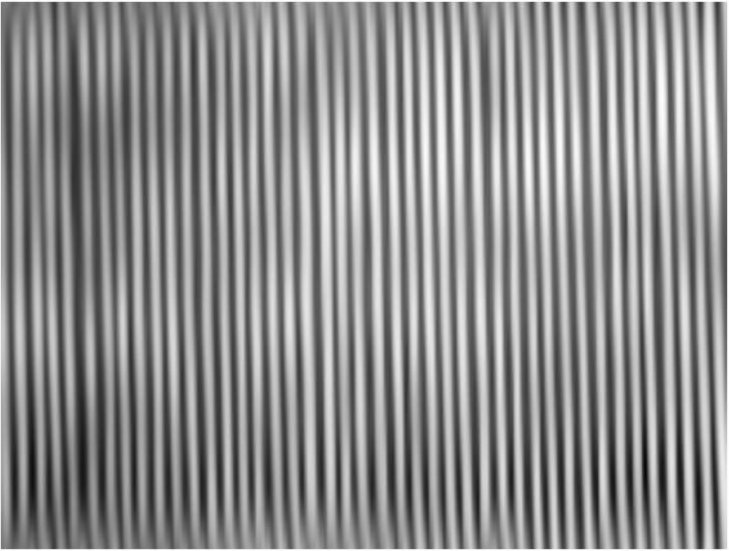

Gambar-11 Gambar benang lusi hasil rekonstruksi

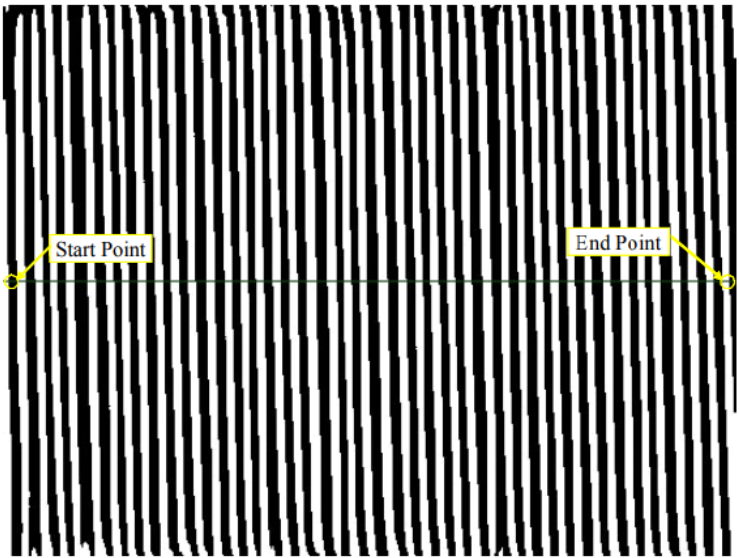

Gambar-12 Gambar benang lusi hasil proses *threshold*

*Pixel* hitam pada Gambar 6 mewakili benang lungsin sementara piksel putih menunjukkan celah. Untuk menghitung jumlah benang, garis horizontal gambar dipilih sebagai garis standar pengukuran seperti yang ditunjukkan pada Gambar 6. *Pixel* pada garis standar dilalui dari kiri ke kanan. Jumlah benang sepanjang garis standar kemudian dihitung.

Dimensi awal pada garis standar termasuk pada benang batas atau interstis. *Pixel* yang memiliki nilai abu-abu yang berlawanan dengan piksel pertama (dari kiri) pada baris standar dipilih sebagai titik awal. Ini berarti bahwa jika piksel pertama adalah titik putih, piksel hitam pertama selama traversal dipilih sebagai titik awal untuk pengukuran kerapatan, dan sebaliknya.

*Pixel* yang termasuk keseluruhan benang atau daerah silangan terakhir yang berlawanan dengan titik awal dipilih sebagai titik akhir. Misalnya, pada Gambar 6, titik awal adalah piksel hitam yang dipilih sebagai piksel awal dari benang pertama di sebelah kiri, yang diberi label sebagai "Start point" pada gambar. Pixel putih yang tergabung dalam silangan terakhir di sebelah kanan gambar dipilih sebagai titik akhir, diberi label sebagai "End Point".

Setelah mendapatkan titik awal dan titik akhir pada citra biner, kerapatan dapat diukur dengan menghitung jumlah benang selama melintasi garis standar. Pada garis standar, hanya ada dua jenis sinyal abu-abu: 0 dan 1. Bila sinyalnya bergantian, piksel saat ini adalah titik alternatif dari benang dan interstis. Jika nilai abu-abu dari titik alternatif sama dengan titik awal, benang baru dianggap ditemukan. Berdasarkan metode ini, semua benang bisa ditemukan.

Setelah mendapatkan jumlah benang, dengan posisi titik awal dan titik akhir, kerapatan benang lusi dapat dihitung dengan pembesaran gambar yang diambil. Metode perhitungannya dilakukan sebagai berikut

$$D_{warp} = \frac{N}{(EP - SP + 1) \, x \, Scale} \quad (12)$$

dimana $D_{warp}$ adalah kerapatan benang lusi (helai/cm), $N$ adalah jumlah benang lusi, $EP$ adalah koordinat titik akhir, $SP$ adalah koordinat titik awal, dan *Scale* adalah nilai pembesaran dari gambar kain yang diambil dalam (cm / pixel)

Dalam Gambar-12, $N$ adalah 42 benang, $EP$ 506 dan $SP$ 4, dan *Scale* 0,002363 cm / pixel. $D_{warp}$ dihitung dengan hasil 35,3 benang / cm.

Kepadatan benang pakan dapat diukur dengan metode yang serupa. Benang-benang pakan yang dipisahkan dari benang lusi direkonstruksi dengan sinyal frekuensi yang dipilih pada Gambar-13a, dan gambar setelah pengolahan *threshold* ditunjukkan pada Gambar-13b. Jumlah benang pakan dihitung dan kerapatan pakan dihitung secara otomatis sesuai dengan persamaan $D_{warp} = \frac{N}{(EP-SP+1) \, x \, Scale}$. Kerapatan pakan dari Gambar-13 adalah 27,0 benang / cm.

Dalam sistem pengukuran, kerapatan benang pakan diperiksa dengan memutar sampel sejauh 90°. Kepadatan benang lungsin dan benang pakan diukur pada pembesaran sepuluh kali lipat. Nilai rata-rata dihitung secara otomatis dan digunakan sebagai hasil pengukuran akhir.

Tabel-3 Perbandingan hasil pengukuran kerapatan benang pada kain dengan cara otomatis (pengolah citra) dan cara manual (dengan kaca pembesar)

| Fabric sample | Warp density, threads/cm | | Weft density, threads/cm | | Measurement error, % | |
|---|---|---|---|---|---|---|
| | automatic | manual | automatic | manual | warp | weft |
| P1 | 53.7 | 53.5 | 37.2 | 37.0 | 0.37 | 0.54 |
| P2 | 56.5 | 56.5 | 38.5 | 38.5 | 0.00 | 0.00 |
| P3 | 61.1 | 61.0 | 39.0 | 39.0 | 0.16 | 0.00 |
| P4 | 41.3 | 41.5 | 18.1 | 18.0 | 0.48 | 0.56 |
| P5 | 65.6 | 65.5 | 39.3 | 39.5 | 0.15 | 0.51 |
| P6 | 42.5 | 42.5 | 30.9 | 31.0 | 0.00 | 0.32 |
| P7 | 47.2 | 47.0 | 21.6 | 21.5 | 0.43 | 0.47 |
| T1 | 61.8 | 62.0 | 27.3 | 27.5 | 0.32 | 0.73 |
| T2 | 48.4 | 48.5 | 36.3 | 36.0 | 0.21 | 0.83 |
| T3 | 65.6 | 65.5 | 35.7 | 35.5 | 0.15 | 0.56 |
| T4 | 50.2 | 50.0 | 21.5 | 21.5 | 0.40 | 0.00 |
| T5 | 54.2 | 54.0 | 20.7 | 20.5 | 0.37 | 0.98 |
| S1 | 99.6 | 99.5 | 78.3 | 78.5 | 0.10 | 0.25 |
| S2 | 76.2 | 76.0 | 42.4 | 42.5 | 0.26 | 0.24 |
| S3 | 87.9 | 88.0 | 44.7 | 44.5 | 0.11 | 0.45 |

Kain yang digunakan dalam percobaan semuanya diukur secara otomatis dengan metode pengolahan digital. Hasil inspeksi untuk kerapatan benang lusi dan benang pakan pada penelitian yang dilakukan oleh Pan dkk (2015) dicantumkan pada Tabel-3. Pengukuran kepadatan kain manual ini adalah pengukuran secara manual dengan bantuan kaca pembesar.

$Error$ pengukuran dihitung untuk mengevaluasi hasil pemeriksaan metode otomatis dan manual. $Error$ dihitung dengan persamaan berikut ini

$$Error\% = \left|\frac{D_A - D_M}{D_M}\right| x 100\% \tag{13}$$

dimana, $D_A$ adalah hasil pengukuran kerapatan benang dengan menggunakan pengolah citra secara otomatis, $D_M$ adalah kerapatan yang diperoleh dengan metoda pengukuran kerapatan benang manual.

Dapat dilihat bahwa hasil pengukuran otomatis dan manual bernilai hampir sama, dengan kesalahan maksimum 0,98% (kerapatan lusi dari sampel kain T5). Kesalahan ini terutama berasal dari daerah pengukuran yang berbeda (variasi kerapatan pada daerah tertentu pada kain). Dari hasil tersebut dapat disimpulkan bahwa deteksi kerapatan otomatis dapat digunakan untuk mengganti metode deteksi kerapatan manual saat ini. Sistem yang digunakan oleh Pan dkk (2015) dapat dikatakan dapat dijadikan suatu metode alternatif pengukuran kerapatan kain.

Dengan sistem pengukuran Pan dkk (2015), kerapatan benang kain tenunan dapat diperiksa secara otomatis. Citra rekonstruksi benang digunakan untuk membantu operator menentukan orientasi kain (menentukan arah lusi dan pakan). Operator hanya perlu memilih sepuluh daerah berbeda di kain untuk mendapatkan hasil pengukuran akhir dari kerapatan kain. Kepadatan sampel kain dapat diperiksa dalam satu menit dengan sistem ini, yang jauh lebih cepat dan lebih mudah daripada metode manual tradisional.

## 4. DETEKSI CACAT KAIN MENGGUNAKAN IMAGE PROCESSING

Di industri tekstil, deteksi cacat kain memegang peranan penting dalam pengendalian mutu. Nadaf dkk (2017) mengatakan bahwa deteksi cacat atau pemeriksaan adalah proses mengidentifikasi dan menemukan cacat pada kain. Industri tekstil sangat memperhatikan kualitas dalam proses produksinya. Hal ini bertujuan untuk menghasilkan barang dengan kualitas terbaik dalam waktu sesingkat mungkin (Nadaf dkk, 2017). Inspeksi mutu merupakan aspek penting dalam industri manufaktur. Kualitas kain bisa diperbaiki dengan mengurangi cacat pada kain. Pegangan pada kain bergantung pada struktur kain. Pegangan pada kain dapat dikatakan kasar, lembut, halus, lembut, berkilau, dan lain-lain. Tekstur atau pegangan kain dapat berbeda tergantung pada jenis tenunan yang digunakan. Tekstur dapat berbeda untuk semua jenis kain, katun, sutra, wol, kulit, dan juga linen.

Terdapat beberapa jenis cacat yang sering terjadi pada manufaktur pembuatan kain, yaitu cacat *floating*, cacat *pin mark*, cacat *stain*, cacat *slub*, cacat *ladder*, cacat *hole* atau lubang. Contoh kenampakan cacat tersebut dapat dilihat pada Gambar-13, Gambar-14, Gambar-15, Gambar-16, Gambar-17 dan Gambar-18.

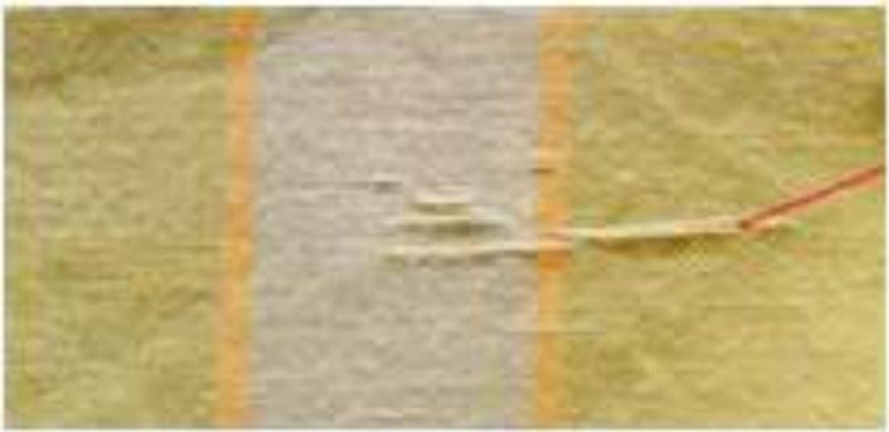

Gambar-13 Cacat *float*

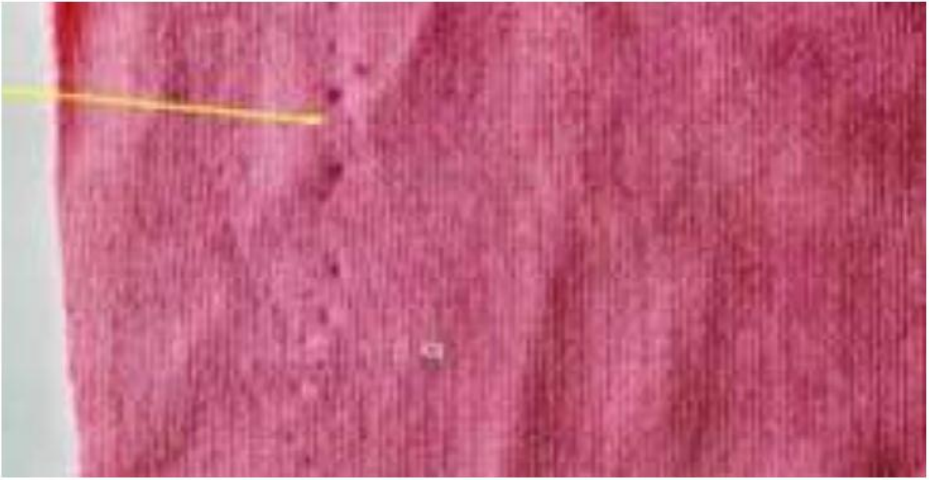

Gambar-14 Cacat *pin mark*

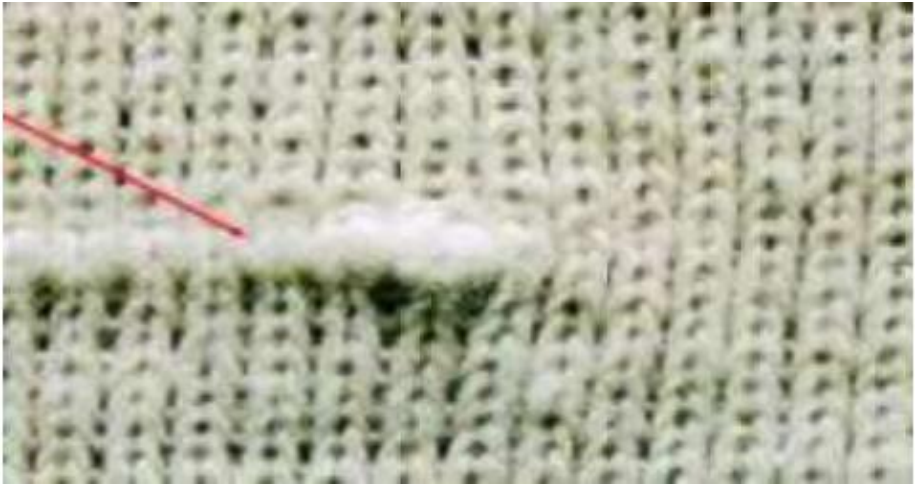

Gambar-15 Cacat *stain*

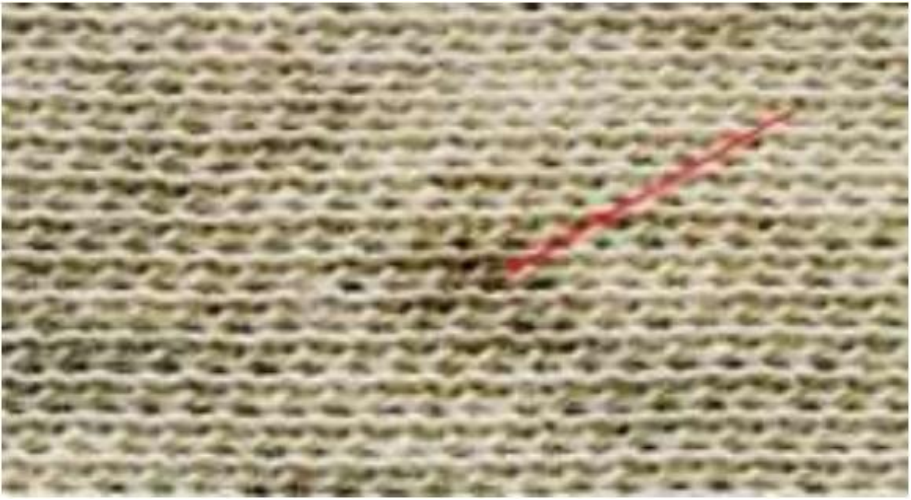

Gambar-16 Cacat *slub*

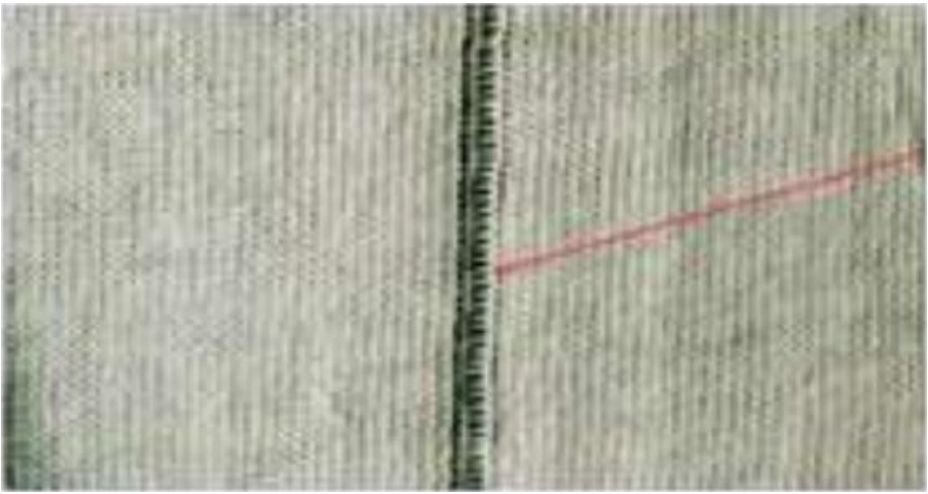

Gambar-17 Cacat *ladder*

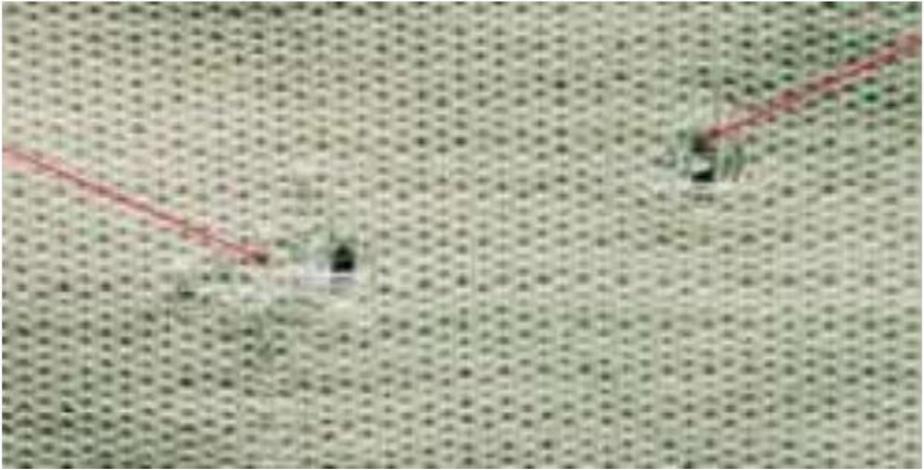

Gambar-18 Cacat *hole* (lubang)

Proses pra-pengolahan citra memegang peranan penting pada penelitian Nadaf dkk (2017). Nadaf dkk (2017) menggunakan sistem pendeteksian yang bersifat *real-time* untuk nendeteksi cacat pada kain. Proses pra-pengolahan citra yang dilakukan oleh Nadaf dkk (2017) adalah sebagai berikut:

1. Penangkapan citra (*image capturing*)
2. Pengubahan citra Gray ke citra RGB
3. Pemerataan histogram (*histogram equalization*)
4. Pengubahan menjadi citra Biner
5. Operasi *edge detection*
6. Operasi *feature extraction*

Pada proses ini, telah digunakan sebuah perangkat kamera untuk mengambil citra pada daerah kain. Citra yang ditangkap disimpan pada sebuah perangkat komputer. Gambar yang ditangkap kemudian diproses dengan aplikasi MATLAB.

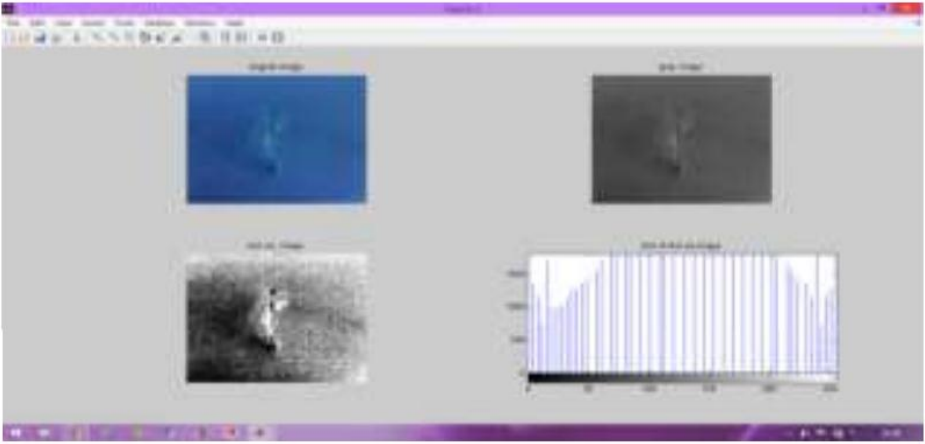

Gambar-19 Hasil pra-pengolahan citra (Nadaf dkk, 2017)

Untuk melakukan proses *edge detection* pada citra yang diolah, telah dilakukan proses *edge detection* dengan metode Canny. Adapun metode lain yang dapat digunakan untuk proses *edge detection*, yaitu dapat dilakukan dengan metode Sobel, metode Canny, dan metode Prewitt.

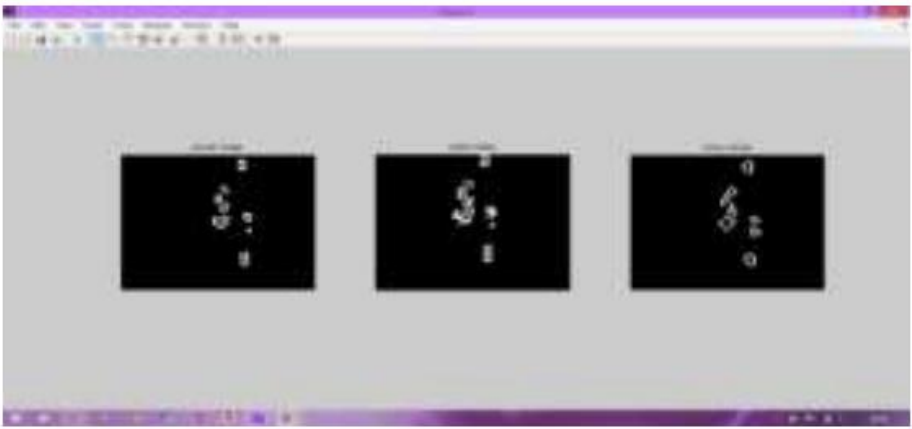

Gambar-20 Hasil proses operasi *edge detection*

Proses pengolahan citra digital yang telah dilakukan dengan menggunakan aplikasi MATLAB. Berdasarkan diagram alir tersebut, citra digital yang telah selesai dilakukan proses pra-pengolahan akan dilakukan proses *edge detection.* Hasil proses deteksi akan masuk keadalam sebuah *controlling system.*

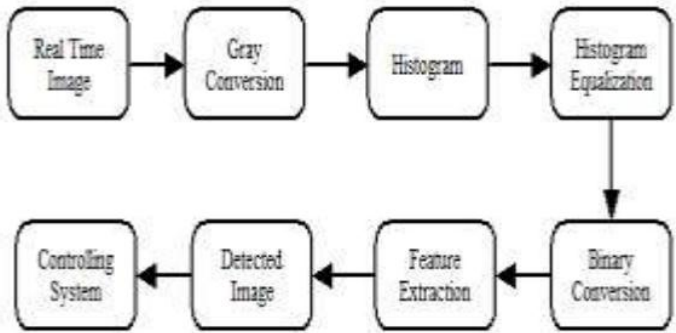

Gambar-21 Diagram alir proses pengolahan citra yang dilakukan dengan aplikasi MATLAB

Menurut Nadaf dkk (2017), sistem pendeteksian yang telah dibuatnya memiliki kelebihan sebagai berikut:

1. Dengan mengaplikasikan proses ini, akan lebih mudah untuk mengindentifikasi cacat pada kain berdasarkan citra digital.
2. Metoda tersebut dapat memberikan tingkat akurasi yang lebih tinggi dibandingkan dengan metode yang lama.
3. Metode identifikasi cacat dengan menggunakan *image processing* dapat memberikan hasil dengan waktu yang lebih cepat dibandingka dengan metode yang lama.
4. Mengurangi jumlah pekerja yang harus digunakan untuk proses inspeksi.

Adapun kelemahan dari alat yang telah dirancang oleh Nadaf dkk (2017), yaitu

1. Masih membutuhkan keterampilan pekerja untuk mengecek kinerja metode baru ini

2. Kesalahan dapat saja terjadi pada perangkat lunak pengolah citra
3. Apabila listrik terputus pada sistem, kemungkinan terjadinya *loss data* dan kesalahan deteksi cacat pada kain.

Nadaf dkk (2017) mengemukakan bahwa ada beberapa hal yang dapat dikembangkan dari metode ini, yaitu

1. Di industri tekstil kita bisa mendeteksi kesalahan pada citra asli & bisa menghilangkan kesalahan dengan menggunakan sistem kontrol tingkat lanjut.
2. Cacat pada industri tekstil dapat dideteksi dan diakses menggunakan fitur serial komunikasi wireless
3. Dalam penelitian tersebut hanya menggunakan software MATLAB, namun nantinya bisa digunakan software baru seperti SCILAB, Virtual LAB & Computer Vision.

Selain penelitian Nadaf dkk (2017), adapula penelitian yang telah dilakukan oleh Das dkk (2016), yaitu penelitian untuk mendeteksi cacat kain pada kain jute. Pada penelitian tersebut, Das skk (2017) mengklasifikasikan 9 cacat pada kain jute, yaitu *harness breakdown, warp blur, mispick, warp float, foreign fiber, color fly, oil spot, water damage* dan *knot* (kenampakan cacat dapat dilihat pada halaman 287-290). Selama ini, inspeksi cacat pada kain tenun jute dilakukan dengan cara manual. Menurut Das, dkk (2017), inspeksi cacat secara manual terbagi menjadi dua jenis, yaitu dengan cara tradisional dan modern.

Inspeksi manual tradisional dilakukan oleh seorang inspector kain yang memiliki pengalaman yang cukup untuk mengamati dan melakukan penilaian pada bahan tekstil. Inspeksi tersebut dilakukan dengan menggunakan perangkat lup (kaca pembesar) untuk mengamati struktur pada kain, serta untuk menggolongkan setiap cacat pada kain.

Pada metode inspeksi manual modern, proses inspeksi dibantu dengan sebuah perangkat meja inspeksi untuk mengamati dan menghitung jumlah cacat pada kain. Metode ini juga banyak dikenal dengan nama kegiatan *grading* kain. *Grade* kain diklasifikasikan dengan menghitung jumlah cacat pada sebuah kain.

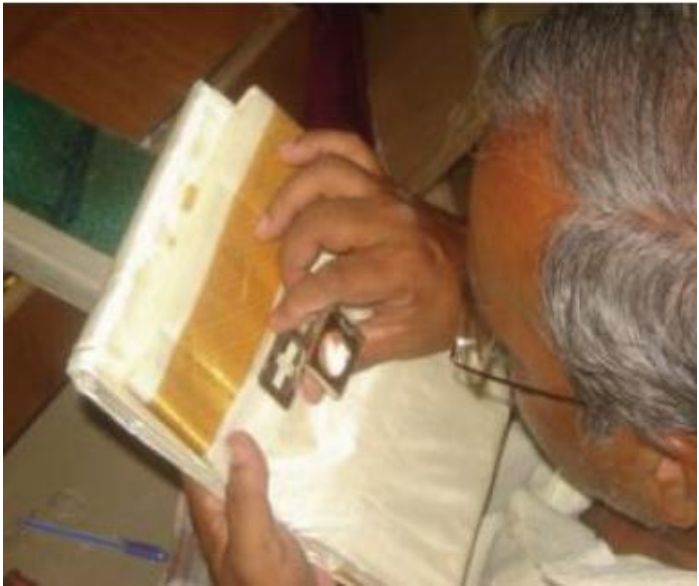
Gambar-32 Inspeksi cacat kain jute manual tradisional

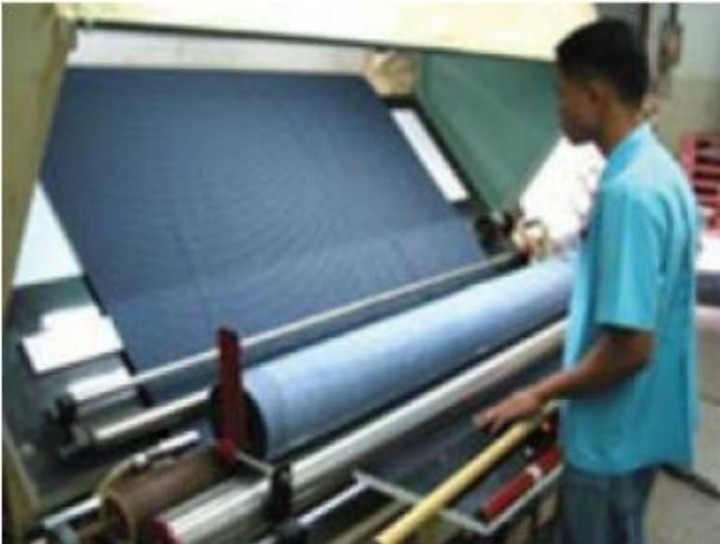
Gambar-33 Inspeksi cacat kain jute manual modern

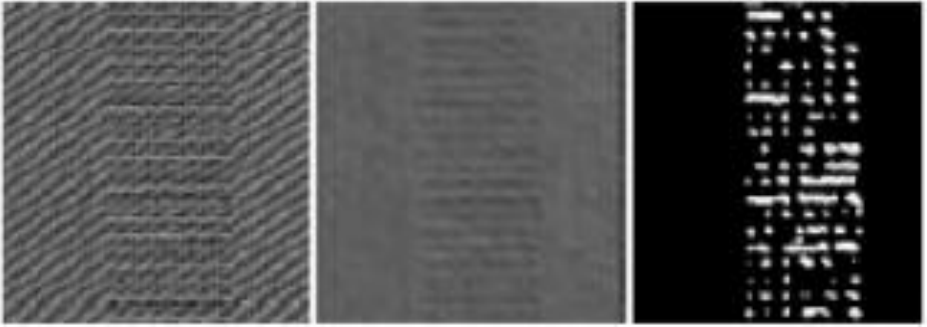

Gambar-22 Cacat *harness breakdown*

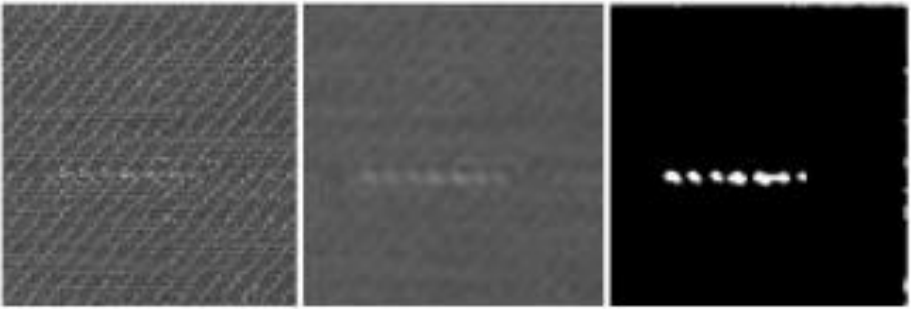

Gambar-23 Cacat *warp blur*

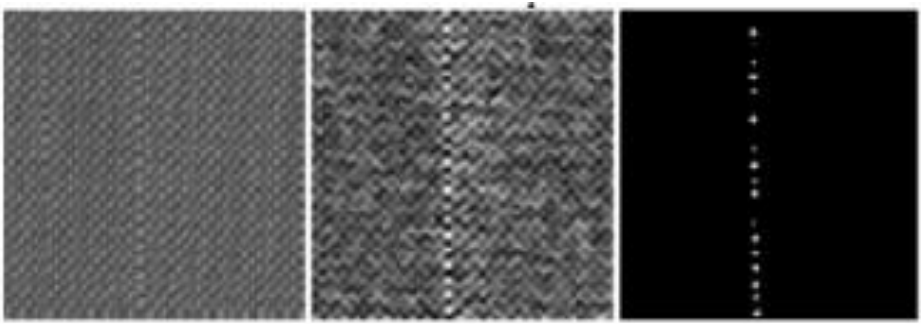

Gambar-24 Cacat *mispick*

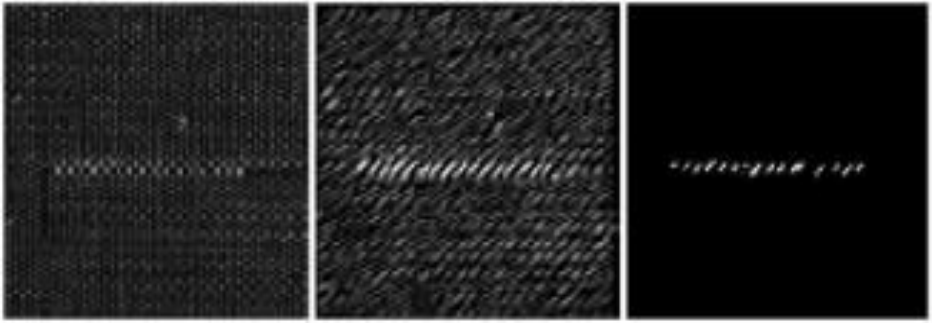

Gambar-25 Cacat *warp float*

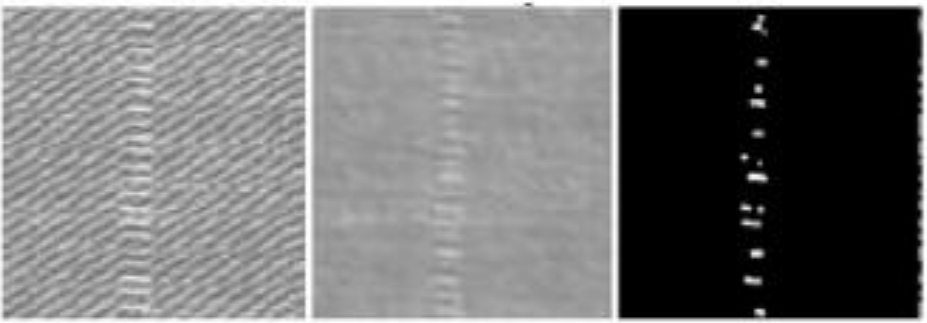

Gambar-26 Cacat *mispick*

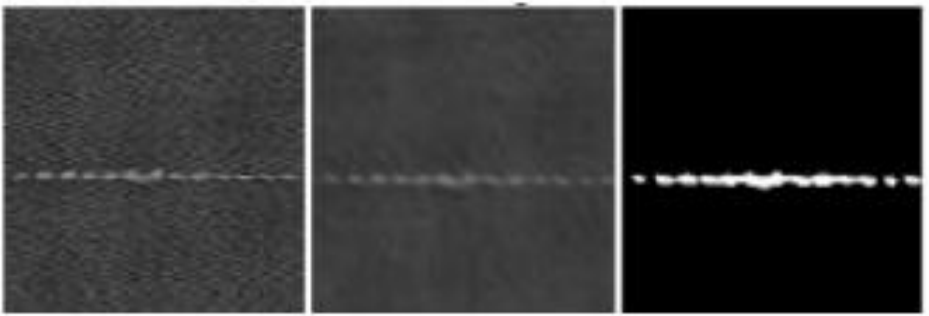

Gambar-27 Cacat *foreign fiber*

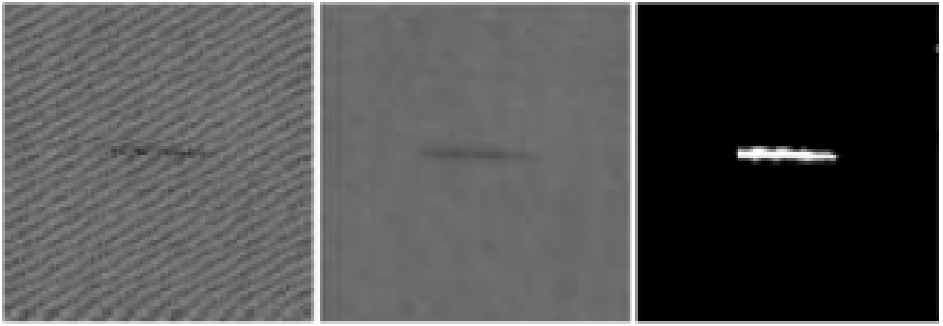

Gambar-28 Cacat *color fly*

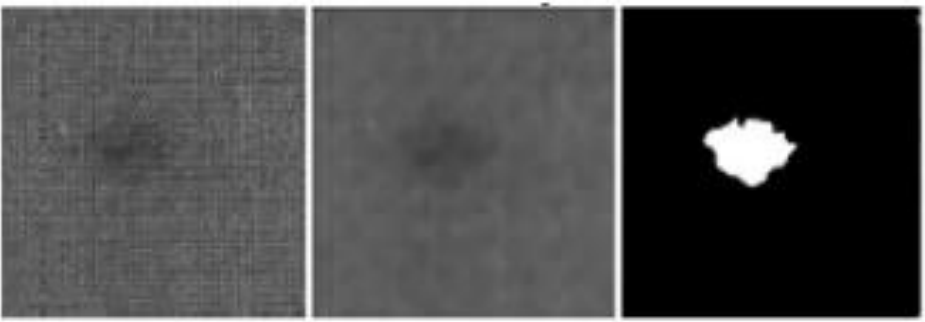

Gambar-29 Cacat *oil spot*

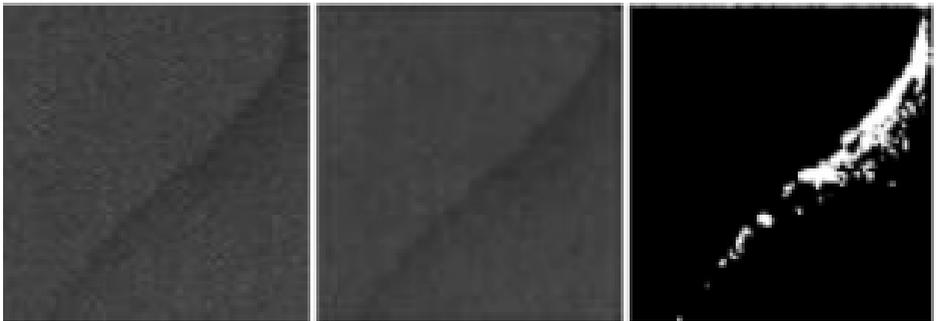

Gambar-30 Cacat *water damage*

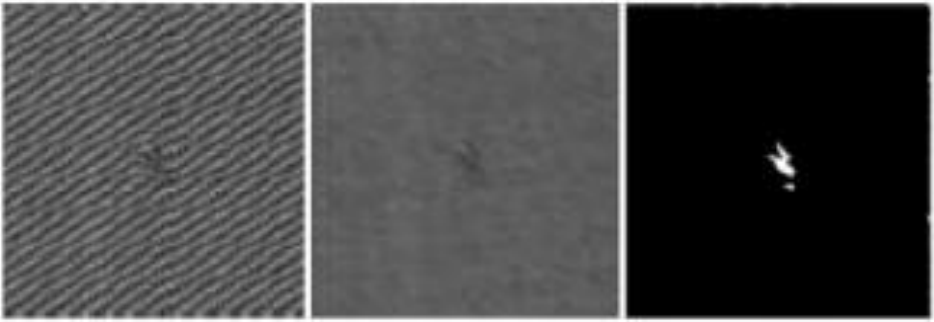

Gambar-31 Cacat *knot*

Gambar-gambar tersebut merupakan hasil pengolahan dari aplikasi pengolah citra digital untuk setiap jenis cacat pada kain jute. Tahapan proses pengolahan citra yang dilakukan oleh Das dkk (2016) adalah penangkapan citra, proses pra-pengelolaan citra, proses *feature extraction*, proses segmentasi citra, proses pengolahan citra tingkat tinggi dan proses pengambilang keputusan.

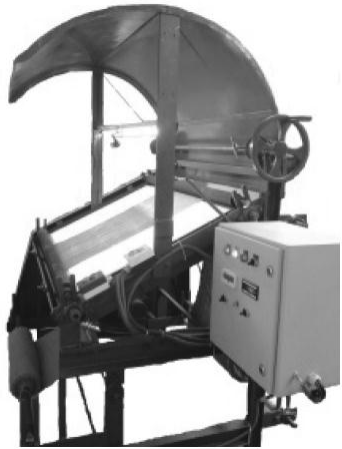

Gambar-32 Meja inspeksi hasil modifikasi Das dkk (2016)

Das dkk (2016) telah merancang sebuah perangkat meja inspeksi yang dimodifikasi untuk dapat menghasilkan suatu tangkapan citra pada permukaan kain. Gambar-32 memperlihatkan perangkat meja inspeksi yang telah dimodifikasi.

Meja pemeriksaan kain telah dikembangkan untuk pemeriksaan kain secara online dengan menggunakan kamera untuk menganalisis cacat pada kain. Kain melewati platform semi transparan dengan cahaya yang dapat disesuaikan sesuai visibilitas pada kain (citra harus memiliki pencahayaan seragam dan citra yang jernih). Perangkat tersebut terdiri dari rol kain, rol pengulur kain dan rol lainnya untuk mengatur pergerakan kain. Gerakan rol ini dioperasikan oleh motor yang dapat dikontrol dengan saklar ke belakang, tombol start-stop dan tombol kontrol kecepatan. Rol tesebut memiliki sensor untuk menampilkan panjang kain dan kecepatan kain pada platform. Terdapat pula tombol reset untuk mengatur ulang counter panjang kain. Kotak panel juga berisi lampu atas, lampu bawah dan tombol lampu. Perangkat tersebut juga memiliki stand kamera horizontal yang bisa bergerak naik turun secara vertikal dengan roda pengatur. Kamera sudah terpasang di stand kamera diatas meja pemeriksaan. Kamera telah terintegrasi ke perangkat lunak inspeksi menggunakan kabel USB dan menangkap data online gambar kain. Semua penyesuaian kamera seperti fokus, kecerahan, kontras, dan lain-lain telah dikontrol dari perangkat lunak inspeksi. Persiapan kain jute dan pengambilan gambar pada platform inspeksi menggunakan perangkat lunak inspeksi telah dilakukan. Perangkat lunak inspeksi mengukur dan menganalisis cacat kain dan dinyatakan sebagai no. cacat, peta cacat, titik & intensitas grafik cacat pada posisi kain yang berbeda, panjang cacat yang ditangkap, dll. Meja inspeksi telah dibuat oleh Das skk (2016) memiliki spesifikasi berikut:

1. Dapat mengulur kain pada mode jalan dan mode *inching*
2. Dapat mengulur ke arah maju dan mundur
3. Kecepatan penguluran pada kain dapat diubah
4. Memiliki lebar 46 inci
5. Memiliki penutup atas untuk menghindari pengaruh interferensi cahaya yang berlebihan dari arah lampu
6. Memiliki kamera yang dapat diatur ketinggiannya

7. Memiliki *counter* kecepatan penguluran
8. Memiliki *counter* panjang kain

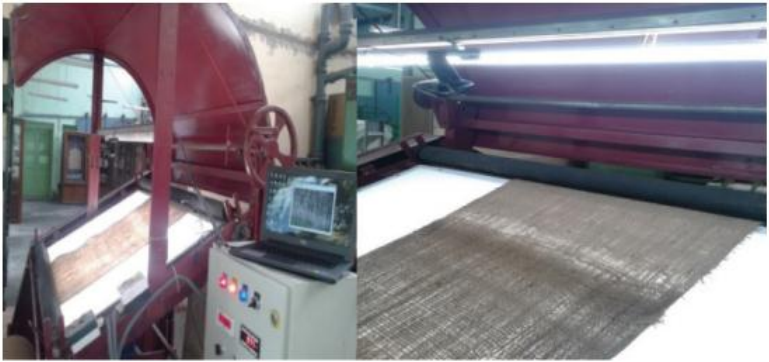

Gambar-33 Proses pendeteksian cacat kain dengan alat yang dirancang Das dkk (2016)

Setelah selesai melakukan inspeksi pada kain, pekerjaan pengembangan perangkat lunak dimulai. Untuk pengembangan perangkat lunak standar, jenis video kain yang berbeda telah ditangkap pada kecepatan kain yang berbeda dan ketinggian yang berbeda. Pengolahan citra awal telah dilakukan pada citra tunggal. Segmentasi semua area cacat dari keseluruhan citra telah dilakukan dalam citra tunggal dengan menggunakan teknik pengolahan citra. Kemudian dihitung total area gambar, total semua area cacat dan perhitungan persentase area yang cacat. Sebuah tanda indikasi telah ditampilkan di semua area yang cacat pada setiap sentroid area cacat untuk melihat kekurangan kain secara jelas. Ada dua jenis indikasi indikator merah dan biru. Indikator merah menunjukkan daerah kental dan indikator biru menunjukkan area kain yang tipis. Pengolahan citra telah dilakukan di semua frame video untuk deteksi cacat waktu nyata dan menghitung persentase kerusakan. Perangkat lunak ini telah menampilkan citra *real time* kain pada meja inspeksi. Perangkat lunak ini memiliki tombol start and stop untuk menangkap video dan analisis. Sebelum memulai pengambilan video, penyesuaian ukuran cacat minimum diperlukan menggunakan scroll bar. Atas dasar penyesuaian ukuran cacat minimum, perangkat lunak akan mengabaikan ukuran cacat dibawah ukuran kecil tersebut. Setelah menekan tombol start, perangkat lunak akan memulai mendeteksi cacat dan menghitung persentase cacat kain bergerak pada meja inspeksi secara real time. Setelah menekan tombol stop, panjang kain yang ditangkap telah dihitung berdasarkan jumlah roller yang ditampilkan pada panel. Setelah menekan tombol grafik, perangkat lunak telah menampilkan grafik konsentrasi cacat dan grafik titik cacat pada berbagai panjang kain. Berbagai percobaan telah dilakukan pada berbagai jenis kain (gramase yang berbeda) dan kecepatan yang berbeda. Komputer telah menghitung cacat deteksi dan persentase cacat pada kecepatan pergerakan kain sekitar empat meter per menit.

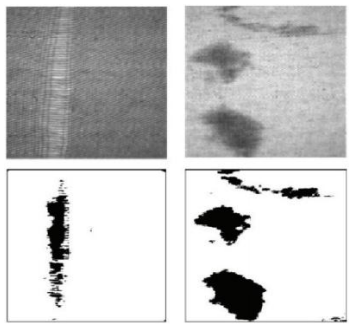

Gambar-34 Hasil segmenasi area cacat pada kain

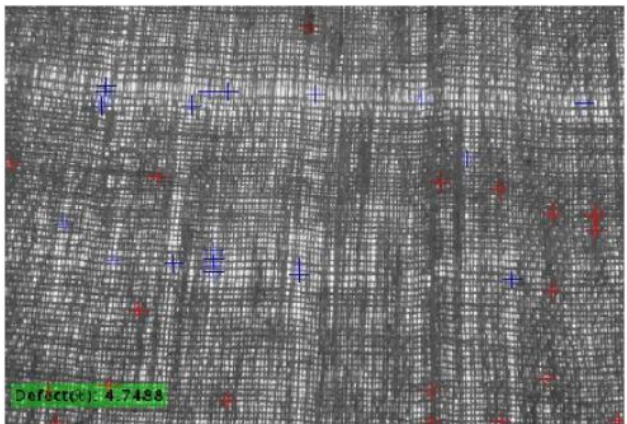

Gambar-35 Sistem pendeteksian cacat *real-time*

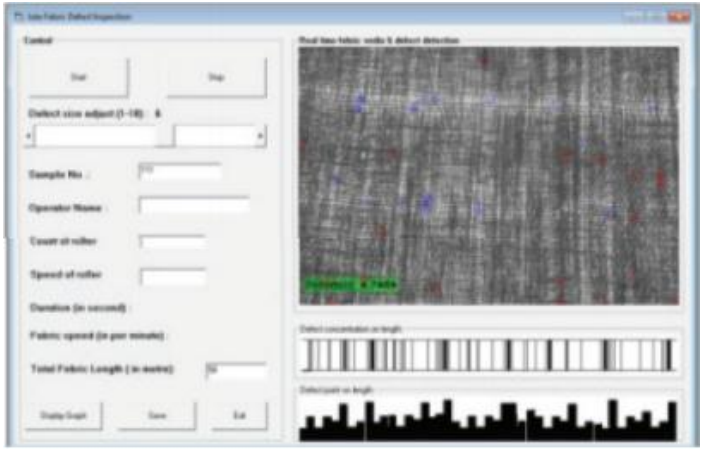
Gambar-36 Tampilan aplikasi pendeteksi cacat

## 5. REFERENCES

[1] A. Kumar, "Automated defect defection in textured materials," Ph.D. dissertation, Dept. Elect.Electron. Eng., Univ. Hong Kong, Hong Kong, May 2001
[2] B.Karunamoorthy, Dr.D.Somasundareswari, S.P.Sethu "AUTOMATED PATTERNED FABRIC FAULT DETECTION USING IMAGE PROCESSING TECHNIQUE INMATLAB".InternationalJournal of Advanced Research in Computer Engineering & Technology (IJARCET) .Volume 4. Issue 1, January 2015
[3] Berlin J, Worley S, Ramey H. Measuring the Cross-Sectional Area of Cotton Fibres with an Image Analyzer. Textile Res. J. 1981; 51: 109-113.
[4] Bhanumati, P., Nasira, G.M. Fabric inspection system using Artificial Neural Network, International Journal of Computer Engineering
[5] C.H. Chan, G.K.H. Pang, "Fabric defect detection by Fourier analysis", IEEE Trans. Industry Applications vol.36, no.5 (Sep/Oct 2000) 1267–1276.
[6] Cybulska M. Assessing Yarn Structure with Image Analysis Methods. Textile Res. J. 1999; 69: 369-373.

[7] Cybulska M. Analysis of Warp Destruction in the Process of Weaving Using the System for Assessment of the Yarn Structure. Fibres & Textiles in Eastern Europe 1997; 5(4): 68–72.

[8] Das S, Sengupta S, Shambhu V.B, Ray D.P. 2016. Defect detection of jute fabric using image processing. Economic Affairs 61(2): 273-280.

[9] Escofet J, Millán M,Ралló M. Modeling of woven fabric structures based on Fourier image analysis. Journal of Applied Optics 2001; 40, 34: 6171-6176.

[10] Feng Y, Li L. Automatic measurement of weave count with wavelet transfer. J. Text. Res. 2001; 22, 2: 94-95.

[11] H. Sari-Sarraf and J. S. Goddard, "On-line optical measurement and monitoring of yarn densityinwovenfabrics,"Proc.SPIE,vol.2899,pp.44-452,1996.

[12] He F, Li L, Xu J. Woven fabric density measure based on adaptive wavelets transform. J. Text. Res. 2007; 28, 2: 32-35.

[13] Huang X, Bresee R. Characterizing Nonwoven Web Structure Using Image Analysis Techniques. INDA 1993; 5: 13-211.

[14] J. G. Campbell, C. Fraley, F. Murtagh, and A. E. Raftery, "Linear flaw detection in woven textiles using model-based clustering," Dept. Statistics, Univ. Washington, Seattle, WA, Tech. Rep. 314, Jul. 1996, pp. 1–15.

[15] Jasińska I. Assessment of a Fabric Surface after the Pilling Process Based on Image Analysis. FIBRES & TEXTILES in Eastern Europe 2009; 17, 2 (73): 55-58.

[16] Jeong YJ, Jang J. Applying image analysis to automatic inspection of fabric density for woven fabrics. Fibers and Polymers 2005; 26, 2: 156-161.

[17] Kang TJ, Chang HK, Kung WO. Automatic recognition of fabric weave patterns by digital image analysis. Textile Research Journal 1999; 69, 2: 77-83

[18] Kumar, "Computer-Vision-Based Fabric Defect Detection: A Survey," IEEE Trans. Ind. Electron., vol. 55, no. I, pp. 348-363, Jan. 2008.

[19] Kuo, CFJ, Shih CY, Ho CE, Peng KC. Application of computer vision in the automatic identification and classification of woven fabric weave patterns. Textile Research Journal 2010; 80, 20: 2144-2157.

[20] Lachkar A, Gadi T, Benslimane R, D'Oraziob L, Martuscellib E. Textile woven-fabric recognition by using Fourier image-analysis techniques. Part I: A fully automatic approach for crossed- points detection. J. Text. Inst. 2003; 94, 3: 194-201.

[21] Li L, Chen X, Huang X. Automatic inspection of weaving density for woven fabrics using adaptive wavelets. J. Donghua Univ. Natural Science Edition 2005; 31, 1: 63-66.

[22] Lim J, Kim SM. Analysis of woven fabric structure using image analysis and artificial intelligence. Fibers and Polymers 2011; 12, 8: 1062-1068.

[23] Lin JJ. Applying a co-occurrence matrix to automatic inspection of weaving density for woven fabrics. Textile Research Journal 2002; 72, 6: 486-490.

[24] Liqing L, Tingting Jia, Xia Chen. Automatic recognition of fabric structures based on digital image decomposition. Indian J. of Fibre & Textile Res. 2008; 33, 388-391.

[25] Liu J, Jiang H, Pan R, Gao W, Xu M. Evaluation of yarn evenness in fabric based on image processing. Textile Research Journal 2012; 82, 10: 1026-1037.

[26] Mahajan, P.M., Kolhe, S.R., Patil, P.M. 2009. A review of automatic fabric defect detection techniques, Advances in Computational Research 1: 18-29.

[27] Mahure, J., Kulkarni, Y.C. 2014. Fabric faults processing: perfections and imperfections, International Journal of Computer Networking, Wireless and Mobile Communications (IJCNWMC) 4.

[28] Mak, K.L., Peng, P., Yiu, K.F.C. 2009. Fabric defect detection using morphological filters, International Journal of Emerging Technology and Advanced Engineering, Elsevier 27: 1585-1592.

[29] Masajtis J. Thread Image Processing in the Estimation of Repetition of Yarn Structure. Fibres & Textiles in Eastern Europe 1997; 10(4): 68-72.

[30] Mirjalili SA, Ekhtiyari E. Wrinkle Assessment of Fabric Using Image Processing. FIBRES & TEXTILES in Eastern Europe 2010; 18, 5 (82): 60-63.
[31] Mourssa A, Dupont D, Steen D, Zeng X. Structure analysis and surface simulation of woven fabrics using fast Fourier transform techniques. Journal of the Textile Institute 2010; 101, 6: 556-570.
[32] Nadaf F.S, Kamble N.P, Gadekar R.B. 2017. Fabric Fault Detection Using Digital Image Processing. International Journal on Recent and Innovation Trends in Computing and Communication ISSN: 2321-8169 Vol 5, pp. 128-130.
[33] Newman, T.S., Jain, A.K. 1995. A survey of automated visual inspection, Computer Vision Image Understanding 61(2): 231–262.
[34] Niblack W. An introduction to digital image processing. Prentice Hall, 1986.
[35] Pan R, Gao W, Liu J, Wang H. Automatic inspection of woven fabric density of solid colour fabric density by the Hough transform. Fibres & Textiles in Eastern Europe2010; 18, 4: 46-51.
[36] Pan R, Gao W, Liu J, Wang H. Automatic recognition of woven fabric pattern based on image processing and BP neural network. Journal of the Textile Institute 2011; 102, 1:19-30. [16] Pan R, Gao W, Liu J, Wang H, Qian, X. Automatic inspection of double-system- melange yarn-dyed fabric density with color-gradient image. Fibers and Polymers, 2011, 12(1): 127-131.
[37] Pierce FT. The Geometry of Cloth Structure. The J. of the Textile Inst. 1937; 28(3): 45-96.
[38] Pohle E. Interlaboratory Test for Wool Fineness Using the PiMc. J. Testing Eval. 1975; 3: 24-26.
[39] Potiyaraj P, Subhakalin C, Sawangharsub B, Udomkichdecha, W. Recognition and re-visualization of woven fabric structures. International Journal of Clothing Science and Technology 2010; 22, 2-3: 79-87.
[40] Priya, S., Kumar, T.A., Paul 2011. V. A novel approach to fabric defect detection using digital image processing, Signal Processing,

Communication, Computing and Networking Technologies (ICSCCN), 2011 International Conference on IEEE 228-232.
[41] Prof. P. Y. Kumbhar, TejaswiniMathpati, RohiniKamaraddiand NamrataKshirsagar. "Textile Fabric Defects Detection and Sorting Using Image Processing".InternationalJournal For Research in Emerging Science and Technology. volume-3, issue-3, mar-2016:ISSN 2349-7610.
[42] Rafael, C. Gonzalez and Richard, E. 2008. Woods, Digital Image Processing, 2nd ed. Prentice Hall Upper Saddle River, New Jersey.
[43] Ralló M, Escofet J, Millán M. Weave-repeat identification by structural analysis of fabric images. Journal of Applied Optics 2003; 42, 17: 3361-3372.
[44] Rao Ananthavaram, R.K., Srinivasa, Rao O., Krishna PMHM. 2012. Automatic Defect Detection of Patterned Fabric by using RB Method and Independent Component Analysis, International Journal of Computer Applications 39(18).
[45] Sabeenian, R.S., Paramasivam, M.E., Dinesh, P.M. 2011. Detection and Location of Defects in Handloom Cottage Silk Fabrics using MRMRFM & MRCSF". International Journal of Technology and Engineering System (IJTES) 2(2).
[46] Semnani D, Hasani H, Behtaj S, Ghorbani E. Surface Roughness Measurement of Weft Knitted Fabrics Using Image Processing. FIBRES & TEXTILES in Eastern Europe 2011; 19; 3 (86): 55-59.
[47] Sezgin M, Sankur B. Survey over image thresholding techniques and quantitative performance evaluation. Journal of Electronic Imaging 2003; 13 (1): 146–165.
[48] Shady E, Abouiiana M, Youssef, S, Gowayed Y, Pastore C. Detection and classification of defects in knitted fabric structures. Textile Res. J. 2006; 76(4): 295-300.
[49] Sun Y, Chen X, Wang X. Automatic recognition of the density of woven fabrics. J. Donghua Univ. Natural Science Edition) 2006; 32, 2: 83-88.
[50] Tamnun, M.E., Fajrana, Z.E., Ahmed, R.I. 2008. Fabric Defect Inspection System Using Neural Network and Microcontroller, Journal of Theoretical and Applied Information Technology.

[51] TaunAnh Pham, "Optimization of texture feature extraction algorithm" master thesis, Delft University of Technology, 2010.
[52] Texture Feature Extraction:Digital image processing by S jayaraman in MGH
[53] Thibodeaux D, Evans J. Cotton Fibre Maturity by Image Analysis. Textile Res. J. 1986; 56: 130-139.
[54] Thilepa R.: A Paper on Automatic Fabrics Fault Processing Using Image Processing Technique in MATLAB, Signal & Image Processing: An International Journal (SIPIJ), Vol. 1, No. 2, December, 2010, pp. 88 - 99.
[55] Tunák M, Linka A, Volf P. Automatic assessing and monitoring of weaving density. FiberPolym. 2009; 10, 6: 830-836
[56] Watanabe A, Kurosaki S, Konoda F. Analysis of Blend Irregularity in Yarns Using Image Processing. Textile Res. J. 1992; 62: 729-735.
[57] Wu Y, Pourdeyhimi B, Spivak M. Texture Evaluation of Carpets Using Image Analysis. Textile Res. J.1991; 61(7): 407-419.
[58] Xu B. Identifying fabric structure with fast Fourier transform techniques. Text. Res. J. 1996; 66, 8: 496-506.
[59] Yu X, Xin B, Gerrge B, Hu J. Fourieranalysis based satin fabric density and weaving pattern extraction. Research Journal of Textile and Apparel 2007;11, 1: 71-80.
[60] Zhang XY, Pan RR, Liu JH, Gao WD, Xu WB. Design Gabor filters in the frequency domain for unsupervised fabric defect detection. Industria Textila 2011; 62, 4: 177-182
[61] Zhang YF, Bresee RR. Fabric Defect Detection and Classification Using Image Analysis. Textile Res. J. 1995; 65: 1-9.
[62] Zhao W, Johnson N, Willard A. Investigating Wool Fibre Damage by Image Analysis. Textile Res. J. 1986; 56: 464-466.
[63] Żurek W, Krucińska I, Adrian H. Distribution of Component Fibres on the Surface of Blend Yarns. Textile Res. J. 1982; 52: 473-478.